\documentclass[prd,tightenlines,%
superscriptaddress,nofootinbib,notitlepage,preprintnumbers]{revtex4}

\usepackage{hyperref}
\usepackage{amsmath}
\usepackage{amssymb}
\usepackage{graphicx}
\usepackage{bm}
\usepackage{amsthm}
\usepackage{cleveref}
\usepackage{indentfirst}

\newcommand{\dd}{\mathrm{d}}
\newcommand{\bea}{\begin{eqnarray}}
\newcommand{\eea}{\end{eqnarray}}
\newcommand{\beq}{\begin{equation}}
\newcommand{\eeq}{\end{equation}}
\newcommand{\normconst}{{\cal C}}
\newcommand{\Y}{X_\xi}
\newcommand{\x}{x}
\newcommand{\y}{y}

\begin{document}

\title{Domain walls without a potential}

\author{C{\'e}dric Deffayet}
\email{cedric.deffayet@iap.fr}
\affiliation{$\mathcal{G}\mathbb{R}\varepsilon{\mathbb{C}}\mathcal{O}$, Institut d'Astrophysique de Paris,\\ UMR 7095, CNRS, Sorbonne Universit{\'e},\\ 98\textsuperscript{bis} boulevard Arago, 75014 Paris, France}
\affiliation{IHES, Le Bois-Marie, 35 route de Chartres, F-91440 Bures-sur-Yvette, France}
\author{Fran\c{c}ois Larrouturou}
\email{francois.larrouturou@iap.fr}
\affiliation{$\mathcal{G}\mathbb{R}\varepsilon{\mathbb{C}}\mathcal{O}$, Institut d'Astrophysique de Paris,\\ UMR 7095, CNRS, Sorbonne Universit{\'e},\\ 98\textsuperscript{bis} boulevard Arago, 75014 Paris, France}

\begin{abstract}
We show that domain walls, or kinks, can be constructed in simple scalar theories where the scalar has no potential. These theories belong to a class of k-essence where the Lagrangian vanishes identically when one lets the derivatives of the scalar vanish. The domain walls we construct have positive energy and stable quadratic perturbations. As particular cases, we find families of theories with domain walls and their quadratic perturbations identical to the ones of the canonical Mexican hat or sine-Gordon scalar theories. We show that canonical and non canonical cases are   
nevertheless distinguishable via higher order perturbations or a careful examination of the energies. In particular, in contrast to the usual case, our walls are local minima of the energy among the field configuration having some fixed topological charge, but not global minima. 
\end{abstract}

\date{\today}

\maketitle


\section{Introduction}
\label{sec:intro}

Topological and non-topological solitons play an important role in various domains of physics ranging from liquid crystals, fluid mechanics to cosmology (see e.g. \cite{Vilenkin:2000jqa,Vachaspati:2006zz,Weinberg:2012pjx,Shnir:2018yzp,Manton:2004tk}). The simplest and canonical example of such objects are certainly domain walls, or kinks, which are known to exist in particular in simple scalar theories where the vacuum manifold possesses several connected components. Considering such a theory, with a scalar $\phi$, and a potential $V(\phi)$, domain walls can exist if the potential has more than one minimum. The purpose of this work is to show that similar domain wall solutions exist in scalar theories with no potential; {\it i.e.} theories where the Lagrangian vanishes identically 
when the derivatives of the scalar vanish. Among such theories, we will concentrate here on Lorentz invariant theories where the Lagrangian depends both on the real scalar field $\phi$ and on its the kinetic term $X$, defined by 
\beq \label{defX}
X = - \frac{1}{2} \eta^{\mu\nu}\partial_\mu\phi \partial_\nu\phi,
\eeq  assuming space-time is endowed with a Lorentzian flat metric $\eta_{\mu \nu}$ (we will not consider here gravitating solutions). Hence we will consider Lagrangians $\mathcal{L}$  of the form 
\begin{equation} \label{defLagrangian}
\mathcal{L} = P\left(\phi,X\right)
\end{equation}
where the dependence of $P$ on $X$ and $\phi$ is non trivial and in particular not given by a sum of a free kinetic energy $X$ and potential energy $V(\phi)$. Such theories have been considered in many instances and are usually denoted as k-essence in the context of cosmology and gravitation \cite{Bekenstein:1984tv,ArmendarizPicon:2000ah,ArmendarizPicon:2000dh,ArmendarizPicon:1999rj}. They have second order equations of motion and can even be generalized to Lagrangian including up to second derivatives of the field, the so-called Horndeski theories \cite{Horndeski:1974wa,Deffayet:2011gz}. Such theories can be used in particular to mimic dark matter via the MOND paradigm \cite{Bekenstein:1984tv,Milgrom:1983ca} or even possibly as dark matter itself \cite{ArmendarizPicon:2005nz}, to generate inflation without a potential \cite{ArmendarizPicon:1999rj,Garriga:1999vw,ArmendarizPicon:2000yi} or get a late time accelerated expansion \cite{ArmendarizPicon:2000dh,ArkaniHamed:2003uy}.

In this context, the possibility of finding solitonic configurations in theories with non-canonical kinetic terms was considered in several works, in particular in the Horndeski framework \cite{Babichev06,Babichev:2007tn,Sarangi:2007mj,BLMO07,Jin:2007fz,ASW09,AKSW09,BLM15,CT16,ZGFL18,Brax:2003rs,Brax:2003wf,Babichev:2008qv,Andrews:2010eh,Endlich:2010zj,Carrillo-Gonzalez:2016yxq,Andrade:2018afh,Andrade:2019ggz} and the corresponding field configuration are sometimes dubbed "k-defects"\cite{Babichev06}. Similar solutions also arose in the past in other contexts for example in the well known Skyrme model~\cite{ZB86}.
The k-defects, in particular, were found to behave differently from standard defects due to the different nature of the kinetic terms \cite{Babichev06,Andrews:2010eh}, however, at least in the single field case, all the existing k-defects, 
 are, despite their name, supported by a non trivial potential in the action, just as the usual topological defects are. {\it I.e.} in the solutions considered so far, $ P\left(\phi,X=0\right)$ has a non trivial dependence in the field. 

Here we show that defects field configurations, specifically kinks, can be obtained in theories with no potential, i.e. theories where the Lagrangian vanishes identically if the kinetic term $X$ is set to zero. This might not come as a surprise considering that one is allowed to freely choose the function $P$ to produce a given specified field profile, however, we will also show that the quadratic perturbation theory around these solutions can be made stable. In fact we will further show that simple models can be considered where both the kink solution and its perturbations are {\it identical} to those of the canonical theories usually considered. We will not attempt here a full classification of the theories allowing such kinks "without a potential" but will only exhibit some simple models as an existence proof and discuss some of the properties of these kinks in comparison with the usual ones. 

This work is organized as follows: in the next section \ref{candwprop} we recall some properties of kinks of usual scalar theories. We then introduce k-essence domain walls (section \ref{seckessenceDW}) and show how one can obtain kinks which have a profile just identical to the one of the canonical mexican hat model  and discuss their stability and topological properties in a non perturbative way. This is then generalized to other canonical profiles including the one of the sine-Gordon model (section \ref{mimickDW}). In a following section, we discuss the perturbation theory around our wall solutions (section \ref{pertsection}) before concluding (section \ref{sec:concl}). Two appendices give technical details on some results introduced in the body of the text.

\section{Canonical domain walls revisited} \label{candwprop}
\subsection{Actions and field equations for canonical domain walls}
Canonical domain walls can be constructed in a fairly standard theory for a scalar field $\phi$ with a Lagrangian of the form 
\begin{equation}\label{eq:P_can_genV}
{\cal L}_\text{can}(\phi,X) = X - V(\phi)\,.
\end{equation}
where the field is assumed to live in a D dimensional flat space-time with metric $\eta_{\mu \nu} = \text{diag}\left(-1,1,\cdots,1\right)$, and $V(\phi)$ is the potential energy.
 In the canonical case, $V$ is chosen so that it has two or more minima (with the same values of the potential $V$) at different values $\phi_{min}^k$ of the field (where $k$ index the different minima).  Domain walls\footnote{Note that we will later specialize to $D=2$ where one calls usually domains wall, kinks. As our result can be easily extended from ``kinks'' in 2 dimensions to ``domain walls'' in arbitrary $D$ we will use both terms interchangeably.} are then obtained as static vacuum solutions $\phi(z)$ of the field equations which only depend on one space-like direction $z$ (to simplify the discussion, one also usually assumes that the field live in $D=2$ dimensions) and interpolate between different adjacent minima $\phi_{min}^{-\infty}$ at $z=-\infty$ and $\phi_{min}^{+\infty}$ at $z=+\infty$. 
For the canonical models (\ref{eq:P_can_genV}), a given vacuum profile  $\phi(z)$ obeys the vacuum field equation which has the first integral 
\bea \label{FirstInt}
\frac{1}{2} \phi'^2- V  \approx {\cal J}_0,
\eea 
where ${\cal J}_0$ is a constant, and here and henceforth a prime means a derivative w.r.t. $z$. Note further that, when we want to stress that a given expression is valid only on shell for the background domain wall solution, we will replace there the 
straight symbols (\emph{eg.} "$=$") (designating off-shell relations) by curly symbols (\emph{eg.} "$\approx$"). As a consequence, the kink profile obeys 
\bea
\phi' \approx \pm \sqrt{2(V+{\cal J}_0)}
\eea

\subsection{Some energy considerations}
A standard trick due to Bogomolny \cite{Bogomolny:1975de} (that we write here in a slightly non standard way) allows then to discuss easily the total energy\footnote{Throughout this work, we use the same letter ${\cal H}$ to denote the total energy and the energy density of the field configuration, the difference between the two is just indicated 
by the dependence on $z$ of the energy density which is explicitly indicated when necessary.} ${\cal H}$ of such a configuration. Indeed, this energy (or the energy per unit transverse to the direction $z$ if $D >2$) is given by
by the integral over $z$ of the Hamiltonian density ${\cal H}(z)$ given by 
\bea
{\cal H}(z) = \frac{1}{2}\phi'^2(z) + V(\phi), \label{Hofzcan}
\eea
so that one has 
\bea
{\cal H} &=& \int {\cal H}(z) dz \\
&=& \int dz \left[ \frac{1}{2} \left(\phi' \pm \sqrt{2 (V +{\cal J}_1)}\right)^2 \mp \sqrt{2(V+{\cal J}_1)}\phi'- {\cal J}_1\right] \label{Ebis} \\
&\geq & \mp \int dz  \left[\sqrt{2(V+{\cal J}_1)} \phi' \pm {\cal J}_1\right] \label{Bogolbound}
\eea
where ${\cal J}_1$ is an arbitrary constant. 
Choosing ${\cal J}_1={\cal J}_0$ we see that the last bound is saturated for a solution of the field equations obeying (\ref{FirstInt}), as the square appearing in the right hand side of 
(\ref{Ebis}) vanishes. Moreover, it is possible to make this energy finite for such a solution representing a domain wall. In this case, one takes ${\cal J}_0 =0$ and the domain wall energy ${\cal H}_{dw}$ is given by the simple expression 
\bea \label{DWE}
{\cal H}_{dw} &=& \pm \int dz  \left[\sqrt{2V} \phi' \right] = \pm \int_{\phi_{min}^{-\infty}}^{\phi_{min}^{+\infty}} d\phi \sqrt{2V}
\eea
We will later enforce this finiteness as well as demand that the  energy density of the wall is locally finite. Thus we shall require that 
\bea
\int {\cal H}(z) dz < + \infty \label{mathcalHtot} \\
\forall z, \: \left\vert\mathcal{H}(z)\right\vert < + \infty \label{mathcalH0}
\eea

\subsection{Changing variables}
The simple form of the first integral (\ref{FirstInt}) can be used to enlighten the nature of the canonical domain wall solutions as well as ease the finding of the solutions to be discussed thereafter. Indeed, for a generic $\phi$, we define $\psi$ as obeying 
\beq \label{defpsi}
d \psi = \pm \frac{d \phi}{\sqrt{2 V}}. 
\eeq 
I.e., the $\phi(\psi)$ solution of the above equation is given by the same functional dependance as $\phi(z)$ solution of the domain wall profile equation (\ref{FirstInt}) with ${\cal J}_0 = 0 $.
And using the new variable $\psi$ as field variable, the domain wall field equation simply read $\psi'= 1$, and in the $\psi$ variable, the solution is then simply represented by\footnote{\label{zorigin} Note that here and henceforth one can freely choose the position of the domain wall. For simplicity, we will hence assume it lays at the origin $z=0$.}  $\psi = z$.
Using the $\psi$ variable, we see that the Lagrangian (\ref{eq:P_can_genV}) simply reads 
\bea
{\cal L}_\text{can}(\phi,X) = 2 v(\psi)\left(X_\psi -\frac{1}{2}\right)&\equiv& {\cal L}_\text{can}(\psi,X_\psi)\\
&\equiv& v(\psi) w(X_\psi) 
\eea 
where $v(\psi)$ is defined simply by the relation $v(\psi) = V(\phi(\psi))$, $X_\psi$ is defined as in (\ref{defX}) replacing there $\phi$ by $\psi$, and the above equation also defines the function $w(X_\psi)$. 
Considering the above Lagrangian as a starting point, and looking for a one dimensional profile $\psi(z)$, we see that the part of the field equations deriving from this Lagrangian and not proportional to second derivatives of the field simply reads
\bea
v'(\psi)\left(2 X_\psi w'(X_\psi)- w(X_\psi)\right) =0 
\eea 
Hence, looking for a profile of the form $\psi= \lambda z$, and using that for such a profile one has obviously $\psi''= 0$ and $X_\psi \approx -\lambda ^2/2 $, we see that we get a solution provided $-\lambda^2/2 $ is a root of the function  $y$ defined by 
\bea
y(X_\psi) = 2 X_\psi w'(X_\psi)- w(X_\psi).
\eea
In the canonical case, one has $w(X_\psi) = 2 X_\psi - 1$ and hence $y(X_\psi)= 2 X_\psi + 1$. Obviously $\lambda=\pm1$ generates a solution irrespectively of the form of $v$ (say provided that $v$ does not vanish as $\psi$ varies over the real line).
To get a proper domain wall, one should then check that the obtained profile has localized energy and is stable. 
The previous expression (\ref{DWE}) yield the following form of the energy density 
\bea
{\cal H}_{dw} (z) = 2 v\left(\psi(z)\right)
\eea
yielding the total energy
\bea
{\cal H}_{dw} &=& 2 \int_{-\infty}^{+\infty}  v(\psi) d \psi.
\eea
Hence, a necessary condition to have a domain wall is that the above integral converges.

\subsection{Some canonical models} \label{somecanonicalmodel}
Among the most studied and well known cases which have these properties is the model with the mexican hat potential 
\begin{equation} \label{mhpot}
V_{mh} = \frac{1}{2}\left(1-\phi^2\right)^2 
\end{equation} 
The kink and antikink solutions are  given by the profiles
\begin{equation} \label{mhprofile}
\phi_{mh}(z) = \pm \tanh(z)
\end{equation}
and interpolate between the vaccua $\phi^{\pm \infty}_{min} = \pm 1$. This also yields the following relation between $\phi$ and $\psi$ as defined in equation (\ref{defpsi}) 
\begin{equation} \label{phipsimh}
\phi = \tanh(\psi) \Leftrightarrow \psi = \tanh^{-1} \phi.
\end{equation}
Using the variable $\psi$, the Lagrangian reads 
\bea \label{lmhpsi}
{\cal L}_\text{mh}(\psi,X_\psi) = \frac{X_\psi - \frac{1}{2}}{\cosh^4(\psi)},\\
\eea
the function $v(\psi)$ is given here by  
\bea
v_{mh}(\psi) = \left(2 \cosh^{4} \psi\right)^{-1} 
\eea 
and the energy of the solution is just found to be 
\bea \label{mhenergy}
{\cal H}_{mh} &=& \int_{-\infty}^{+\infty}  \frac{d \psi}{\cosh(\psi)^{4}}   = \frac{4}{3}.
\eea
Another case of interest is the sine-Gordon potential 
\begin{equation} \label{msGpot}
V_{sG} = 1-\cos\left(\phi\right)
\end{equation}
which obviously has the infinitely many minima $V=0$ at the fields values $\phi^k_{min} = 2 \pi k$.  
The kink profile which interpolate between the adjacent minima $\phi^k_{min}$ and $\phi^{k+1}_{min}$ is obtained to be 
\begin{equation} \label{sGprofile}
\phi_{sG}(z) = 2 \pi k + 4 \arctan e^{z}.
\end{equation}
Remarkably, the sine-Gordon theory looks very similar to the mexican hat theory (\ref{lmhpsi}) when using the $\psi$ variable. Indeed, in that case, we get that 
the relation between $\psi$ and $\phi$ is given by 
\bea
\phi(\psi) = 2 \pi k + 4 \arctan e^{\psi} \Leftrightarrow \psi = (-1)^k \ln \left|\tan \left(\frac{\phi}{4}\right) \right| 
\eea
and the function $v$ is just obtained to be given by 
\bea
v_{mh}(\psi) = 2  \cosh^{-2} \psi 
\eea
As a result, the sine-Gordon Lagrangian reads now
\bea \label{LsGpsi}
{\cal L}_\text{sG}(\psi,X_\psi) =  \frac{4}{\cosh^2(\psi)}\left(X_\psi - \frac{1}{2}\right).
\eea
Note that of course, the above changes of variables $\phi(\psi)$ are so-defined that its maps the real line (domain of variation of $\psi$) to a finite interval (domain of variation of $\phi$) which does not represent the full range of variation of the $\phi$ field of the original model, and e.g. it does not cover the large values of $\phi$ in the mexican hat potential. Note also that the Lagrangians 
(\ref{lmhpsi}) and (\ref{LsGpsi}) are singular at the end of the interval of definition of $\psi$. We will come back to this issue later. 

\begin{figure}
\centering
\includegraphics[scale=.5]{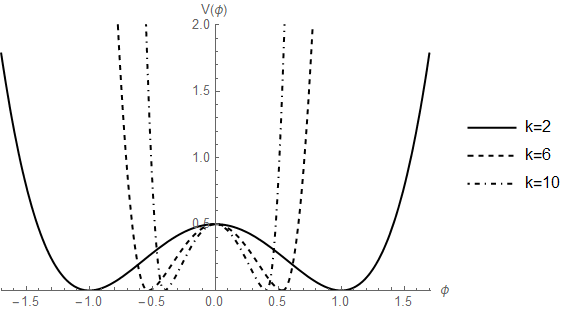}
\caption{Analytic extension of the potential $V(\phi)$ for $k=2$, $k=6$ and $k=10$ respectively.}
\label{mhfig}
\end{figure}

\begin{figure}
\centering
\includegraphics[scale=.5]{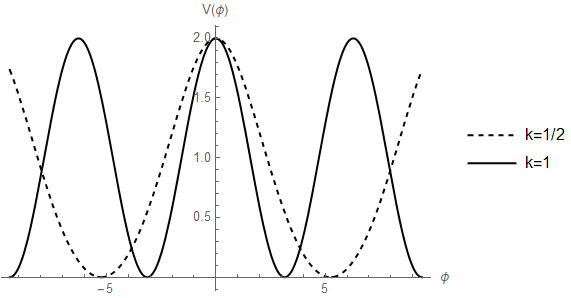}
\caption{Analytic extension of the potential $V(\phi)$ for $k=1$ and $k=1/2$, yielding a periodic profile.}
\label{otherpotfig}
\end{figure}

Given the similarity between Lagrangians (\ref{lmhpsi}) and (\ref{LsGpsi}), we can easily generalize these canonical models to a larger set with Lagrangians of the forms
\bea \label{LKk}
{\cal L}_\text{k,can}(\psi,X_\psi) =  \frac{{\mathcal K}}{\cosh^{2k}(\psi)}\left(X_\psi - \frac{1}{2}\right),
\eea
where ${\mathcal K}$ is some positive constant and $k$ an integer (an even larger family exists letting $k$ be half integer). It is easy to see that $\psi' = \pm 1$  provides a solution of the field equations of the kink type. The energy of this solution is finite and given by 
\bea
{\cal H}_\text{k,can} = {\mathcal K}  \int_{-\infty}^{+\infty} \frac{\mathrm{d} z }{\cosh^{2k} z} = {\mathcal K}\; {\cal I}_{k},
\eea
where ${\cal I}_{k}$ can be computed as 
\bea \label{intek}
{\cal I}_{k}=  \int_{-\infty}^{+\infty} \frac{dz}{\cosh^{2k}(z)} =\frac{\sqrt{\pi}\;\Gamma(k)}{\Gamma(k+1/2)}\;\;\;\; {\rm for} \;\;\; k \geq \frac{1}{2},
\eea
where the above expression holds in particular for integers\footnote{Note that whenever $k$ is a an integer, ${\cal I}_{k}$ can also be expressed as $ 2 \times 4^{k-1} ((k-1)!)^2/(2k-1)!$} and half integers $k$. 
Consider now the change of variable of the form  
\bea
\phi[\psi] &=& \sqrt{\mathcal K} \int_0^\psi\!\frac{\mathrm{d} u}{\cosh^{k} u}.
\eea
When $\psi$ varies over the whole real line, the interval of variation of $\phi$ 
 is just given by $\left]-\frac{\sqrt{\mathcal{K}} \;}{2}{\mathcal I}_{\frac{k}{2}} , \frac{\sqrt{\mathcal{K}}\; }{2}{\mathcal I}_{\frac{k}{2}} \right[$ 
and because $\cosh$ is a positive function, we see that the above defined $\phi[\psi]$ is invertible into a $\psi[\phi]$ on this interval.
This change of variable puts the Lagrangian (\ref{LKk}) in the standard form (\ref{eq:P_can_genV}) with 
the specific potential 
\bea
V \left( \phi \right) \equiv \frac{{\mathcal K}}{2} \cosh^{-2k} \left(\psi[\phi]\right)\;,
\eea
where, at this stage, $V$ is defined for $\phi \in \left]-\frac{\sqrt{\mathcal{K}} \;}{2}{\mathcal I}_{\frac{k}{2}} , \frac{\sqrt{\mathcal{K}}\; }{2}{\mathcal I}_{\frac{k}{2}} \right[$. However, it is easy to see that $dV/d\phi$ vanishes at the ends of this interval (where $\psi$ diverges) allowing to extend the domain of variation of $\phi$ to the entire real line, either by making $V$ periodic (which is always possible, with period then given by $\sqrt{\mathcal{K}} \;{\mathcal I}_{\frac{k}{2}}$) or using an  analytic extension, possibly non periodic. This later possibility arises e.g. in the case of the canonical mexican hat model (\ref{mhpot}), which corresponds to $k=2$. The $k = 6$ or $k = 10$ also yield analytical expressions for $\psi[\phi]$ (however not very enlightening) which in turn result in potentials having a similar shape to the mexican hat one. In turn, the sine-Gordon ($k = 1$) and the $k = 1/2$ cases have potentials which are periodic by analytic extension.
We show these potentials on figures \ref{mhfig} and \ref{otherpotfig}.  
The stability analysis of those models (and their natural generalisation to the k-essence framework) is presented later, in sec \ref{sec:sGdomainwall}.

\subsection{Stability and topology}
The stability of the canonical domain walls can be adressed in several ways. Before recalling in the next subsection some standard results on perturbations of canonical domain walls; we first discuss here their non perturbative stability appealing to some "topological" arguments. We feel that this discussion is often obscured in the literature by an intrication of "topological" and non "topological" arguments and we would like to clarify this below as it matters for the discussion of the stability of non standard domain walls to be introduced later. 

We recall first that the bound (\ref{Bogolbound}) on the total energy also holds for time dependent solutions as the kinetic energy only adds a positive contribution to the right hand side of (\ref{Hofzcan}). More specifically, we can write the conserved total energy ${\cal H}(t)$  of any field configuration $\phi(t,z)$ as 

\bea
{\cal H}(t) 
&=& \int  dz \left[ \frac{1}{2} \dot{\phi}^2+\frac{1}{2} \left(\phi' \pm \sqrt{2 V }\right)^2 \mp \sqrt{2 V}\phi' \right] \label{Bogolboundgen}
\eea
where a dot means a time derivative. Separating the different contributions, we have ${\cal H}(t) = {\cal H}_{kin}(t) + {\cal H}_{grad}(t)+{\cal H}_{\infty}(t)$ where the terms appearing on the right hand side are given by 
\bea
{\cal H}_{kin}(t) &=& \int   \frac{1}{2} \dot{\phi}^2 dz \label{kinBog}\\
{\cal H}_{grad}(t) &=& \int   \frac{1}{2} \left(\phi' \pm \sqrt{2 V }\right)^2 dz \label{gradBog}\\
{\cal H}_{\infty}(t) &=& \int   \mp \sqrt{2 V}\phi' dz \label{topBog}
\eea
Obviously, ${\cal H}_{kin}$ and ${\cal H}_{grad}$ are positive, so any field configuration has a total energy larger than ${\cal H}_{\infty}$ which in turn is only depending on the values of the field at $z=\pm \infty$ and is just given by ${\cal H}_{dw}$ for a canonical domain wall configuration. 

A standard statement is that the canonical domain walls are stable due to the topology of the vacuum manifold. More specifically, the idea is here that a given vacuum of a canonical theory (\ref{eq:P_can_genV}) is obeying $X=0$ and $\phi=\phi_{min}^k$ for some specific $k$ and then is indexed (classically) by the field value $\phi_{min}^k$. 
In order to have a finite energy, a given domain wall solution must lie in vacuum at $z=\pm \infty$, and the values of the field at $\pm \infty$ cannot change continuously while conserving the finite energy of this solution.  This is usually related to the existence of a "topological charge" $Q$ defined from the current 
\begin{equation} \label{currentdef}
J^\mu = \normconst \epsilon^{\mu \nu} \partial_{\nu} \phi
\end{equation}
where $\normconst$ is a proper normalization constant, and $\epsilon^{\mu \nu}$ is the fully antisymmetric Levi-Civita contravariant tensor.
By construction, this current is conserved irrespectively of the field equations and for a generic field configuration $\phi(t,z)$ one has $J^0 = \normconst \phi'$,  The topological (conserved) charge is then defined as 
\begin{equation}
Q = \int_{z=-\infty}^{+\infty} dz J^0(z) = \normconst\left(\phi(+\infty) - \phi(-\infty)\right).
\end{equation}
The domain wall total energy is, as can be seen from (\ref{DWE}), related to $Q$. 
Note however that this argument on stability is not so clear as it may seem and we would like to discuss it below with some details.

First we note that there is some arbitraryness in the definition of the "topological charge". Indeed, the conservation of the current $J^\mu$ as it is defined above is just obviously a trivial consequence of the antisymmetry of $\epsilon$, so that one could have replaced $\phi$ in the right hand side of (\ref{currentdef}) by any function of $\phi$ and obtained a different conserved current and a different associated charge. Given the form of the decomposition (\ref{Bogolboundgen}) an interesting choice of current $\tilde{J}^\mu$ is given 
\bea
\tilde{J}^{\mu} = \tilde{\normconst} \epsilon^{\mu \nu} \partial_{\nu}\left( \int_{\phi_0}^\phi \sqrt{2 V(u)}du\right)
\eea
where $\tilde{\normconst}$ and $\phi_0$ are some constants, implying that $\tilde{J}^{0} =  \tilde{\normconst} \sqrt{2 V(\phi)}\phi'$, so that the conserved charge is now 
\bea
\tilde{Q} = \int_{z=-\infty}^{+\infty} dz \tilde{J}^0(z) = \mp \tilde{\normconst} {\cal H}_{\infty}
\eea
For a generic field configuration  $\phi(t,z)$ one has now a clear {\it identity} between the topological charge $\tilde{Q}$ and ${\cal H}_{\infty}$ while this was not true using the topological charge $Q$, given that in general ${\cal H}_{\infty}$ does not depend only on the difference of field values at $z = \pm \infty$. 
Note that the form of the charge $\tilde{Q}$ is associated with a superpotential $W(\phi)$ defined by $V(\phi) = \frac{1}{2} \left(\frac{d W}{d \phi}\right)^2$ as observed by Bogomolny \cite{Bogomolny:1975de} (see also e.g. \cite{Manton:2004tk}).

Let us then consider the issue of the stability of a given domain wall profile. To that end we consider a given field configuration $\phi(t,z)$ which only differ 
at time $t=t_0$ from some given domain wall profile $\phi_{dw}(z)$ in a bounded region. Obviously, because: (i) the static (and eternal) domain wall solution (given by $\phi_{dw}(t,z)= \phi_{dw}(z) \; \; \forall t$) has vanishing contributions ${\cal H}_{kin}$ and ${\cal H}_{grad}$, (ii) the field profile $\phi(t_0,z)$ and $\phi_{dw}(z)$ are assumed to differ only in a bounded region and hence have the same energy contribution ${\cal H}_{\infty}$ which is conserved, and (iii) the contributions ${\cal H}_{kin}$ and ${\cal H}_{grad}$ are always positive, we see that the domain wall is an absolute mininum of energy for field configurations having the same conserved charge $\tilde{Q}$, and that no localized perturbation of it can change the topological charge $\tilde{Q}$. This shows that the wall configuration is stable, but we stress that this argument is unrelated to the topology of the vacuum manifold, but only relies on the form of the energy (\ref{Bogolboundgen}).

\subsection{Kinks perturbations}
The perturbative stability of the kinks can be checked by deriving the action for the second order perturbations around them, which is also the starting point for the quantization of these perturbations, using the kinks as vacua. By Fourier decomposing a given such perturbation $\varphi$ as 
$\varphi = \sum \varphi_k(z)e^{i\omega_k t}$  one sees that 
each mode then obeys
\begin{equation}\label{eq:SL_pert}
\left(\mathcal{Z}^{zz} \varphi_k'\right)' - \left(\mathcal{Z}^{00}\,\omega_k^2+\mathcal{M}^2\right)\varphi_k = 0\,.
\end{equation}
where $\mathcal{Z}^{00}$, $\mathcal{Z}^{zz}$ and $\mathcal{M}^2$ are $z$- and model-dependent (i.e. depend on the wall profile). 
The above equation is in the Strum-Liouville form\footnote{Note that it can be put in a Schr\"odinger form by redefining $\varphi \rightarrow (-\mathcal{Z}^{00} \mathcal{Z}^{zz})^{1/4} \varphi$ and $dz \rightarrow (-\mathcal{Z}^{00}/\mathcal{Z}^{zz})^{1/2} dz$, see e.g. \cite{CH53}.} and the modes obey an orthogonality relation 
with the measure $dz (-\mathcal{Z}^{00})$ of the form 
(see e.g. \cite{CH53}) 
\begin{equation}
\int\!\dd z\left(-\mathcal{Z}^{00}\right)\varphi_k\varphi_{k'} = 0 \;\;\;{\text{for} }\;\;\; k\neq k'\,.
\end{equation}
One can show that a generic kink always possesses a $zero$ mode (i.e. a solution of the above (\ref{eq:SL_pert}) with $\omega_k=0$) $\varphi_0 \propto \phi'$ associated with the translation of the defect along $z$. In canonical cases discussed here with potentials (\ref{mhpot}) and (\ref{msGpot}), this zero mode is the lowest lying mode of the spectrum and belongs to a discrete part of the spectrum (in the case of potential (\ref{mhpot}), there is another discrete mode) and can be normalized with the above measure, and there is a continuum above (see e.g. \cite{Vachaspati:2006zz,Weinberg:2012pjx}). 
The conditions 
\begin{subequations}\label{eq:conds}
\begin{align}
-\mathcal{Z}^{00}\mathcal{Z}^{zz} > 0 \;,
 \label{mathcalG} 
\\
 0 < 2\int\!\dd z \mathcal{Z}^{00} X < +\infty \;,
 \label{mathcalN} 
\end{align}
\end{subequations}
are fulfilled, indicating stable perturbations.\footnote{At the price of having non canonical perturbations, we could possibly have allowed $\mathcal{Z}^{zz}$ to vanish and still have stable perturbations. We will not consider this possibility here.}  Indeed, the first condition makes sure that the perturbations are free from tachyonic instabilities, while together with the last condition it implies that the perturbations have positive energy and obey an hyperbolic equation. The last condition is also implying that the zero mode $\varphi_0$ has a finite norm (as one has $\varphi_0^2 \propto \phi'^2 = -2 X$). For the canonical models above, we find 
\bea 
\mathcal{Z}^{zz}  = - \mathcal{Z}^{00} &=& 1 \,,\nonumber \\
\mathcal{M}^2 = 2 \left(3 \phi^2 -1\right) &\approx& 6 \tanh^2(z)-2 \,, \label{pertmh}
\eea
for the mexican hat model (\ref{eq:P_can_genV})-(\ref{mhpot}) and
\bea
\mathcal{Z}^{zz}  = - \mathcal{Z}^{00} &=& 1 \,,\nonumber \\
\mathcal{M}^2 = \cos(\phi) &\approx& 2 \tanh^2(z)-1 \,, \label{pertsG}
\eea
for the sine-Gordon model (\ref{eq:P_can_genV})-(\ref{msGpot}). Once again the two different models exhibit similar features. 
Both obey the conditions (\ref{eq:conds}). Note that we can have stable perturbations even if the squared mass $\mathcal{M}^2(z)$ is locally negative. Indeed, this is what happens above around the origin $z=0$.

\section{k-essence domain walls} \label{seckessenceDW}
\subsection{Generic features}
\label{sec:gen_lin_pert}
Starting from a model with a Lagrangian of the form (\ref{defLagrangian}), and restricting ourselves to a 1+1 dimensional space, with metric $\eta_{\mu\nu} = \text{diag}[-1,1]$, we look for a kink solution $\phi(z)$ with stable quadratic perturbations. 
For such a static configuration, the field equations have the first integral\footnote{Where here and henceforth, we denote with a subscript the derivation \emph{wrt.} $\phi$ or $X$ : \emph{eg.} $P_X = \partial P(\phi,X)/\partial X$.} 
\begin{equation}\label{eq:J_def}
\mathcal{J} = 2X\,P_X-P \approx {\cal J}_0\,,
\end{equation}
where ${\cal J}_0$ is a constant. 
This relation is the equivalent of the canonical (\ref{FirstInt}), up to a sign, and it is related to the field equations of the scalar reading 
\bea \label{backeq}
{\cal E} = \phi''\left( P_X + 2 X P_{XX}\right) + P_\phi - 2 X P_{X \phi} \approx 0.
\eea
One has
\bea \label{JEeq}
\mathcal{J}' = -\mathcal{E} \phi'
\eea
which is valid for an arbitrary number of dimensions $D$. Note that in general (i.e. without assuming any special field configuration, so in particular, without assuming that $\phi$ only depends on one coordinate $z$ as for the domain wall case) a Lagrangian (\ref{defLagrangian}) has to obey some conditions in order for the theory to be consistent for arbitrary field configurations. 
These conditions read \cite{ArmendarizPicon:1999rj,Babichev:2007dw,Babichev:2006vx,Bruneton:2007si,Martin_2013,Babichev:2018uiw}  
\bea
0&<& P_X  \label{cond111}\\
0&<& 2 X P_{XX} + P_X \label{cond222}.
\eea 
The first condition above is necessary in order to have a bounded from below Hamiltonian, while the two conditions together lead to hyperbolic equations of motion. In particular note that the 
second one enters as the coefficient of the second derivative in the field in equation (\ref{backeq}). We will come back to these conditions later.
A domain wall being static, it energy density ${\cal H}(z)$ is simply given by the on-shell value of its Lagrangian 
\bea \label{relationHP}
{\cal H}(z) \approx - P\left(\phi(z)\right)
\eea
and in order to have a proper domain wall solution, we shall demand that the energy conditions  (\ref{mathcalHtot}) and (\ref{mathcalH0}) hold.  We will also look for kinks solutions where ${\cal J}_0$ vanishes, as is the case for kinks of canonical models discussed in the previous section.

The perturbations $\varphi(t,z)$  around a given background configuration $\phi(z)$ have a Lagrangian reading at quadratic order 
\begin{equation}
\delta^{(2)}\mathcal{L} = -\frac{1}{2}\left[\mathcal{Z}^{\mu\nu}\partial_\mu\varphi\,\partial_\nu\varphi + \mathcal{M}^2 \varphi^2 \right]\,,
\end{equation}
where the kinetic matrix is diagonal. Its non trivial components and the squared mass term are given by
\begin{equation} \label{defZZM}
\mathcal{Z}^{00} = - P_X\,,
\qquad
\mathcal{Z}^{zz} = \mathcal{J}_X  = 2X P_{XX} + P_X
\qquad \text{and} \qquad
\mathcal{M}^2 = - \mathcal{E}_\phi = \mathcal{J}_{\phi\phi} - \mathcal{J}_{\phi X}\,\phi''\,.
\end{equation}
Following the same path as in the previous section, we Fourier transform a perturbation as $\varphi(t,z) = \sum \varphi_k(z)e^{i\omega_k t}$ so that every Fourier mode obeys equation (\ref{eq:SL_pert}).
As in the canonical case, we can show that there is always a zero-mode. Indeed, differentiating the equation of motion (\ref{backeq}) of the background field with respect to $z$ yields
\begin{equation}
\mathcal{E}' = \left(\mathcal{J}_X \,\phi''\right)'+\left(\mathcal{J}_{\phi X}\,\phi''-\mathcal{J}_{\phi\phi}\right)\phi' 
= \left(\mathcal{Z}^{zz} \phi''\right)' - \mathcal{M}^2\phi' \approx 0,
\end{equation}
so that the zero mode is given by $\varphi_0(z) \propto \phi'(z)$. In order to have stable perturbations (and hence a stable solution) we shall demand that conditions (\ref{eq:conds}) are fulfilled, as in the canonical case. Note in particular that, as $\varphi_0(z) \propto \phi'(z)$, and as we will be looking for theories having the same domain wall profiles as in the canonical theory (e.g. $\phi \propto \tanh(z)$ or $\phi \propto \arctan\,e^z$) this implies that the zero mode has no node, and hence, following a standard argument, is the lowest lying one. 
 In addition, as we have $\mathcal{Z}^{00} = - P_X$, the condition (\ref{mathcalN}), together with the hypothesis that ${\cal J}_0$ vanishes, implies via equation (\ref{eq:J_def}) that the total energy of the wall obtained via (\ref{mathcalHtot}) (and (\ref{relationHP})) is finite and positive. This also shows that whenever ${\cal J}_0$ vanishes, the normalizability of the zero mode implied by condition 
 (\ref{mathcalN}) is just equivalent to having a wall with finite total energy. In fact, as seen from the definitions (\ref{defZZM}), conditions (\ref{eq:conds}) are equivalent on the wall background to conditions (\ref{cond111}) and (\ref{cond222}).

To summarize, in order to find a proper domain wall with stable perturbations (and assuming ${\cal J}_0 =0$, as we shall now do), it is enough to ask that conditions (\ref{eq:conds}) hold, which in turn implies (\ref{mathcalHtot}) and the normalizability of the zero mode. We will also check that (\ref{mathcalH0}) holds. We recall also that we will look for walls in theories with {\it no} potentials, i.e. in theories where the Lagrangian $P(\phi,X)$ vanishes identically at $X=0$.  

\subsection{Stability conditions} 
\label{sec:gen_cons}

Let's apply the conditions \eqref{eq:conds} to a general potential-free $P(\phi,X)$ case. We will assume that the function $P$ can be power expanded into $\sqrt{-X}$ as in 
\begin{equation} \label{classtheo}
P(\phi,X) = \sum_{n\geq2}\alpha_n(\phi)\left(-2X\right)^{n/2}\,,
\end{equation}
where we have set $\alpha_0$ to zero in order to avoid having a potential as well as set $\alpha_1$ to zero as such a term would not contribute to the field equations when the profile depends only on one spatial direction $z$. Hereafter, we will denote the \emph{background} ({\it i.e.}the domain wall) value of $\vert\phi'\vert$ as $f$ 
so that one has $X \approx - f^2/2$ and $f$ is positive  . We will further consider that $f$ is either a constant or a non trivial function of $\phi$, $f(\phi)$ (which is always the case at least implicity if $\phi(z)$ is locally non constant).  Note further that as we consider here spatial profiles, $X$ is negative, hence the chosen minus sign inside the powers appearing on the right hand side of (\ref{classtheo}). In a more general situation, should we want to keep fractional powers in (\ref{classtheo}), we would rather introduce an absolute value of $X$ for terms with odd $n$ in this expansion.

\subsubsection{No domain-walls for $P(X)$ theories}
\label{sec:gen_cons_PX}

Let's first investigate the simplest $P(X)$ case ({\it i.e.} we assume that $P_\phi=0$). In this case, the on-shell conservation equation (\ref{eq:J_def}) 
can easily be integrated to yield (a non vanishing ${\cal J}_0$ would just add below a trivial constant on the right hand side)
\bea
P(X) = P_0 \,\sqrt{-2X}  \approx  P_0 \,\vert\phi'\vert.
\eea
Such a theory does not fall in the class (\ref{classtheo}), as it just has a non vanishing $\alpha_1$, it does not yield domain walls of the kind we are after here and hence will not be further considered.

\subsubsection{Separable theories}
We then focus on "separable" theories, \emph{i.e.} consider 
\begin{equation}\label{eq:conds_P_sep}
P(\phi,X) = \alpha(\phi)\sum_{n\geq2}\beta_n\left(-2X\right)^{n/2}\,,
\end{equation}
where $\{\beta_n\}_{n\geq2}$ is a collection of constant coefficients. 
In this case, one has simply 
\bea \label{Jseparable}
\mathcal{J} &=& \alpha(\phi)\sum_{n\geq2}\beta_n\left(n-1\right)\left(-2X\right)^{n/2} =  \alpha(\phi) \sum_{n\geq2} \beta_n(n-1) f^n \approx 0 \label{polyeqpourf}
\eea
where the last equality holds for the sought for domain wall, as we assumed ${\cal J}_0=0$.
Hence, leaving aside the case of a vanishing $\alpha$ which would make the theory trivial, we must have $f$ a constant $f_0$, root of the polynomial equation (\ref{polyeqpourf}). In this case, the energy density is given by $\mathcal{H} = -\alpha(\phi) \sum\beta_nf_0^n$, so we have to impose that $\alpha$ is regular everywhere (or at least in the domain of variation of $\phi$ for the domain wall profile). The kinetic matrix of the domain wall perturbations is given by 
\begin{equation}
\mathcal{Z}^{00} =
\alpha(\phi) \sum_{n\geq2}n\beta_nf_0^{n-2}
\qquad \text{and} \qquad
\mathcal{Z}^{zz}  =
-\alpha(\phi) \sum_{n\geq2}n(n-1)\beta_nf_0^{n-2}\,,
\end{equation}
thus the conditions (\ref{mathcalG}) and (\ref{mathcalN}) become respectively 

\begin{subequations}
\begin{equation}
 0 
 < \left(\sum_{n\geq2}n\beta_nf_0^n\right)\left(\sum_{m\geq2}m(m-1)\beta_mf_0^m\right)
\end{equation}
\begin{equation}
0
< -\left(\sum_{n\geq2}n \beta_n  f_0^n\right)\int\!\alpha\left(\phi(z)\right)\, \dd z< + \infty\,.
\end{equation}
\end{subequations}
Note that the case of separable theories (\ref{eq:conds_P_sep}) in fact also covers canonical domain walls discussed in the previous section, as the corresponding canonical Lagrangians can be put in the separable form using the variable $\psi$ (equations (\ref{lmhpsi}) and (\ref{LsGpsi})). 
Using this variable, and {\it not} (we stress) $\phi$, one finds indeed that the canonical domain walls are represented by $f=\vert\psi'\vert$  a constant equal to $f_0=1$. 
However, obviously, theories which are separable in $\phi$ variable cannot support domain wall profiles of the type $\phi= \tanh(z)$, as the corresponding $f$ is not constant.
This would not be true, if one would relax the no-potential hypothesis. E.g. the following separable theory 
\begin{equation}
P(\phi,X) = P_0 \,\frac{e^{\lambda\sqrt{-2X}-\lambda(1-\phi^2)}}{1-\lambda\left(1-\phi^2\right)}\,,
\qquad
(\lambda,P_0) \in\  ]0,1[\times\mathbb{R}^-_\star\,,
\end{equation}
which 
behaves as in the $X\rightarrow0$ limit as
\begin{equation}
P(\phi,0) = \frac{P_0\,e^{-\lambda(1-\phi^2)}}{1-\lambda\left(1-\phi^2\right)}\,.
\end{equation}
admits a stable domain wall with a $\phi=\tanh(z)$ profile (with in this case a non vanishing $\mathcal{J}_0 = -P_0$).

\subsubsection{Non-separable theories}
Let us now focus on the more general case of non-separable theories in the class (\ref{classtheo}) and define $n_0\geq 2$ as the smallest integer $n$ for which $\alpha_n$ is non-vanishing.
We can extract $\alpha_{n_0}$ from the first integral ${\cal J}_0 =  \sum (n-1)\alpha_nf^n =0 $. We find 
\begin{equation}\label{eq:P_nsep_alphan0}
\alpha_{n_0} = - \sum_{n>n_0} \frac{n-1}{n_0-1}\,\alpha_nf^{n-n_0}
\qquad \text{and} \qquad
\mathcal{H} = \sum_{n>n_0}\frac{n-n_0}{n_0-1}\,\alpha_n f^n\,,
\end{equation}
where again, we imply here that $f$ can be locally expressed as a function of $\phi$. So the energy constraint  (\ref{mathcalH0}) is satisfied as long as $f$ and $\alpha_n(\phi)$ do not blow up on the relevant range of variation for $\phi$. The kinetic matrix of the perturbations around the wall profile are given by 
\begin{equation}
\mathcal{Z}^{00} =  \sum_{n>n_0} \alpha_n \frac{n_0-n}{n_0-1} f^{n-2}
\quad \text{and} \quad
\mathcal{Z}^{zz}  =\sum_{n>n_0} (n_0-n)(n-1)\,\alpha_nf^{n-2} \,.
\label{kinmatnonsep}
\end{equation}
So the conditions (\ref{mathcalG}) and (\ref{mathcalN}) become respectively 
\begin{subequations}\label{eq:conds_gen_nonsep}
\begin{equation}
0 < \left(-\sum_{n>n_0}\frac{n-n_0}{n_0-1}\,\alpha_nf^n\right)\left(\sum_{m>n_0}(m-n_0)(m-1)\,\alpha_mf^m\right)\,,\label{eq:conds_gen_nonsep_G}
\end{equation}
\begin{equation} \label{normcondn0}
0<\int \left( \sum_{n>n_0}\frac{n-n_0}{n_0-1}\!\,\alpha_n f^n \right)\, \dd z< + \infty\,.
\end{equation}
\end{subequations}
In addition, one has of course to check that condition (\ref{mathcalH0}) holds. To proceed further, we will be looking in the next section for theories admitting walls with identical profiles 
to the one of the canonical mexican hat theory and further show how this can be generalized.

\section{Mimicking canonical domain wall profiles} \label{mimickDW}
\subsection{Static profiles}
\label{sec:concrete}
We look for Lagrangians that can accommodate an hyperbolic tangent domain-wall, $\phi=\tanh(z)$ identical 
to the one of the mexican-hat model (\ref{mhprofile}). The interest of such configuration is three-folded. First, it will make our wall easy to compare with the usual ones, second
the zero-mode $\varphi_0 \propto \phi' = \cosh^{-2}(z)$ is also the fundamental mode, as it bears no node; and third, the background value of $X$  is easily expressed in terms of $\phi$.
Indeed, $f = \vert\phi'\vert$  obeys the functional relation for the domain wall profile (background)
\bea
f(z) \approx 1-\phi^2(z),
\eea
which can be used to simplifying the calculations. With this in mind, we 
can further assume that we can power expand the function $\alpha_n$ as 
\begin{equation}
\alpha_n = \sum_{p\in \mathbb{Z}} \frac{\beta_{n,p}}{2(n-1)} \left(1-\phi^2\right)^p = \sum_{p\in \mathbb{Z}} \frac{\beta_{n,p}}{2(n-1)}f^p
\end{equation}
where $\beta_{n,p}$ are some constants (and the factor $2(n-1)$ is introduced to simplify formulae below). Note that this expansion is even in $z$ (and $\phi$) as $f(z)$ is. We could have added an odd part as well, however, this would drop out of the crucial normalization condition (\ref{normcondn0}) and we will not consider this possibility in this work, as we do not look for exhaustivity here.
The first step if to check the existence of a domain wall solution in the equation of motion, or rather here using the first integral (\ref{eq:J_def}) with ${\cal J}_0=0$. 
Let us first further simplify the setting by considering the case where only 3 coefficients $\beta_{n,p}$ do not vanish above, i.e. consider a Lagrangian $P$ of the form (we will later come back to a more general form)
\begin{equation}
P(\phi,X) = X +
\frac{\beta_{n,p}}{2(n-1)} \left(1-\phi^2\right)^p\left(-2X\right)^{n/2}
+\frac{\beta_{m,q}}{2(m-1)} \left(1-\phi^2\right)^q\left(-2X\right)^{m/2}\,,
\end{equation}
where we have set in addition $\alpha_2 = -1/2$, so that we also have $n_0=2$.
In order to get a finite energy, we must have $n+p>0$ and $m+p>0$ so that the integrals $\int f^{n+p} dz $ and $\int f^{m+p} dz$ converge in equation (\ref{eq:P_nsep_alphan0}).
Note that we have in particular (see equation (\ref{intek})) 
\bea \label{intekbis}
\int_{-\infty}^{+\infty} f^k(z) dz =  {\cal I}_{k}.
\eea
Next, equation (\ref{eq:P_nsep_alphan0}) imposes 
\begin{equation} \label{condpqnm}
\beta_{n,p}= f^{2-n-p}-\beta_{m,q}\,\,f^{m+q-n-p}\,.
\end{equation}
Assuming a non vanishing $\beta_{n,p}$, we get hence that (as $f$ is not a constant here) 
\bea
p=2-n \;\;\;\; {\rm and} \;\;\;\; q=2-m
\eea
together with the relation 
\bea
\beta_{n,2-n} = 1-\beta_{m,2-m},
\eea
so that we are left with a family of theories parametrized by one parameter $\kappa\equiv \beta_{m,2-m}$, with Lagrangians
\begin{equation}\label{eq:P_our_model}
P_{n,m}(\phi,X) = X + \frac{1-\kappa}{2(n-1)}\frac{\left(-2X\right)^{n/2}}{\left(1-\phi^2\right)^{n-2}}+ \frac{\kappa}{2(m-1)}\frac{\left(-2X\right)^{m/2}}{\left(1-\phi^2\right)^{m-2}}\,.
\end{equation}
The conditions (\ref{condpqnm}) ensures that the energy of the solution is finite. Indeed the energy density is found via (\ref{eq:P_nsep_alphan0}) to be 
\bea
\mathcal{H}(z) = \frac{1}{2}\left(\frac{n-2}{n-1}+\frac{(m-n)\kappa}{(n-1)(m-1)}\right)\cosh^{-4}(z)
\eea
which integrates into a total energy 
\bea
\mathcal{H} = \frac{2}{3}\left(\frac{n-2}{n-1}+\frac{(m-n)\kappa}{(n-1)(m-1)}\right).
\eea
Hence we get a strictly positive energy (density) provided that 
\bea
\frac{n-m}{n-2} \kappa < m-1.
\eea
Let us finally check the constraints (\ref{eq:conds})-(\ref{eq:conds_gen_nonsep}).  The coefficients of the kinetic matrix are found via eq.(\ref{kinmatnonsep}) to be independent of $z$ and given by 
\bea
\mathcal{Z}^{00}&=& -\frac{1}{2}\left(\frac{n-2}{n-1}+\frac{(m-n)\kappa}{(n-1)(m-1)}\right) = -\frac{3}{4} {\mathcal H}, \\
\mathcal{Z}^{zz}&=& \frac{2-n+(n-m)\kappa}{2}.
\eea
As expected we see that the positivity and finiteness of the energy is equivalent to the fullfillment of condition (\ref{mathcalN}), so we just need to check that the other condition (\ref{mathcalG}) is satisfied. As $ \mathcal{Z}^{00}$ is strictly negative, this just amount to check that $\mathcal{Z}^{zz}$ is strictly positive, which is implying that 
\bea
1 < \frac{n-m}{n-2} \kappa.
\eea
Hence, at this point, we have shown that the family of Lagrangians (\ref{eq:P_our_model}) 
does accommodate a hyperbolic tangent configuration $\phi = \pm \tanh(z)$ with stable perturbations as long as $n$ and $m$ are two distinct integers and together with $\kappa$ verify the bounds 
\begin{equation}\label{eq:P_our_modelbound}
 n>2, \;\;\;\;\;\; m>2, \;\;\;\;\;\; 1<\frac{n-m}{n-2}\,\kappa < m-1.\;
\end{equation}
Note in particular, that these bounds cannot be satisfied if $\kappa=0$, hence we need at least two non trivial terms of the form $(1-\phi^2)^{2-n} (-2X)^{n/2}$ in the Lagrangian $P(\phi,X)$. 
However, more terms are allowed and we could have considered a larger family with Lagrangians of the form 
\bea \label{moregeneralform}
P = \sum_{n\geq 2}\kappa_n (1-\phi^2)^{2-n} (-2X)^{n/2}
\eea
where $\kappa_n$ are more than three non vanishing properly chosen constants. We will later derive the conditions the $\kappa_n$ must obey. We see here in particular that the mexican potential appear in the explicit form of the functions $\alpha_n(\phi)$. In fact the family (\ref{moregeneralform}) can even be generalized to 
\bea
P = \sum_{n\geq 2} \kappa_n(\phi)\left[ (-2X)^{n/2}-\frac{n-1}{n_0-1}\left(1-\phi^2\right)^{(n-n_0)} (-2X)^{n_0/2}\right]\,,
\eea
where $n_0$ is an integer strictly greater than 1 (in principle, $\kappa_{n_0}(\phi)$ can vanish) and $\{\kappa_n(\phi)\}$ is a collection of functions obeying 
\bea
&  0 < \left(\sum_{n\neq n_0}\frac{n_0-n}{n_0-1}\frac{\kappa_n(\phi)}{\left(1-\phi^2\right)^{2-n}}\right)\left(\sum_{m\neq n_0}(m-n_0)(m-1)\frac{\kappa_m(\phi)}{\left(1-\phi^2\right)^{2-m}}\right)\\
& 0 < \int \left(\sum_{n\neq n_0}\frac{n-n_0}{n_0-1}\left(1-\phi(z)^2\right)^n\kappa_n(\phi(z))\right) d z < + \infty
\eea
In order to recover (\ref{moregeneralform}), it suffices to take $n_0 = 2$ and $\kappa_n(\phi) = (1-\phi^2)^{2-n}\kappa_n $, and the condition on the collection of constants $\{ \kappa_n\}$ discussed below is automatically satisfied provided the above conditions hold.
If the family (\ref{moregeneralform}) is quite simple, it is not the only one to exhibit such features, and another one, inspired by the DBI action, is presented in \cref{sec:app_DBI}.
 
In the rest of this section, we will mostly focus on the on the family (\ref{eq:P_our_model}) and features of its domain wall solution before discussing its perturbations in the next section.

\subsection{Changing variables}
\label{sec:changing_var}
In order to compare our walls to the canonical ones, and better understand their existence, it is instructive to first use the variable $\psi$ presented in a previous section in equation (\ref{defpsi})
 where $V$ is taken to be the mexican hat potential (\ref{mhpot}). Namely we set $\psi = \tanh^{-1}\phi$ so that the wall solution reads $\psi \approx z$ and the Lagrangian (\ref{eq:P_our_model}) reads now 
\begin{equation} \label{Pnmpsi}
P_{n,m}\left(\psi,X_\psi\right) = \frac{1}{\cosh^4\psi}\left(X_\psi + \frac{1-\kappa}{2(n-1)}\left(-2X_\psi\right)^{n/2}+ \frac{\kappa}{2(m-1)}\left(-2X_\psi\right)^{m/2}\right)\,,
\end{equation}
Comparing this form with (\ref{lmhpsi}) we see that the above family of theories and the canonical scalar with a mexican hat potential belong to the same family of theories with Lagrangians of the form 
\bea \label{psigen}
{\cal L} = \left(\sum_{n \in \mathbb{N}}
\kappa_n (-2 X_\psi)^{n/2}\right)\cosh^{-4} \psi,
\eea
where $\kappa_n$ are constants, and in order to avoid issues with fractional powers of negative expressions, we can restrict the discussion to even integers $n$. Note also that the more general form (\ref{moregeneralform}), once rewritten using the $\psi$ variable, reads also as in the above (\ref{psigen}). One difference between our theories (\ref{moregeneralform}) an the canonical one (\ref{lmhpsi}) is of course the presence in (\ref{lmhpsi}) of a pure potential encoded in a non vanishing $\kappa_0$ above. 
A generic theory (\ref{psigen}) falls in the class of separable theories discussed in the previous section and the expression of the first integral ${\cal J}$ reads then as in (\ref{Jseparable})
\bea
{\cal J} = \left(\sum_{n \in \mathbb{N}}
\kappa_n (n-1) (-2 X_\psi)^{n/2}\right)\cosh^{-4} \psi.
\eea
Hence we see that we can get a domain wall solution $\psi = \lambda z$ (leaving for the time being the possibility that $\lambda$ differs from $\pm 1$) provided 
that ${\cal J}_0$ vanishes and that $\lambda$ is a root of the polynomial (using that $-2 X_\psi \approx \lambda^2$) $\lambda \mapsto \sum_{k \in \mathbb{N}}
\kappa_k (k-1) |\lambda|^k$ hence verifies 
\bea \label{constraintkappa}
\sum_{k \in \mathbb{N}}
\kappa_k (k-1) |\lambda|^{k}  = 0. 
\eea
This holds true with $\lambda = \pm 1$ both for the canonical theory (\ref{eq:P_can_genV})-(\ref{mhpot}) which has $\kappa_0 = \kappa_2 = -1/2$ (and the other $\kappa_k$ vanish), and for the 
family (\ref{Pnmpsi}) which has 
\bea
\kappa_2 &=& -\frac{1}{2} \label{kappa2} \\
\kappa_n &=& \frac{1-\kappa}{2(n-1)} \label{kappan}\\
\kappa_m &=& \frac{\kappa}{2(m-1)} \label{kappam}
\eea

One worrysome aspect of the family of theories (\ref{eq:P_our_model}) (or (\ref{Pnmpsi})) is of course the fact that their Lagrangian appear singular at $\phi = \pm 1$, i.e. at the minima of the mexican hat potential (\ref{mhpot}) which are reached at spatial infinity by the domain wall solution $\phi \approx \pm \tanh(z)$. Note first that, as will be shown later, the quadratic Lagrangian for the perturbations around this solution is nowhere singular (including at $z=\pm \infty$) allowing a well defined  perturbation theory  around the "vacuum" represented by the domain wall.  We also note that, once written with the $\psi$ variable, both the canonical model (\ref{eq:P_can_genV})-(\ref{mhpot}) and the models of the family (\ref{eq:P_our_model}) appear singular at $\psi = \pm \infty$ which correspond to the minima $\phi = \pm 1$. However, going back to the $\phi$ variable for the canonical mexican hat model, one gets rid of this singularity. We now show that, similarly, a change of variable can be made in the models (\ref{eq:P_our_model}) (or (\ref{Pnmpsi})) in order to make the Lagrangian everywhere non singular and in fact to extend elegantly the models ``beyond'' $\psi = \pm \infty$ (or $\phi = \pm 1$).   To see this, it is convenient to  define the variable $\xi_p$ by 
\begin{equation}
\mathrm{d}\xi_p \equiv \frac{\mathrm{d}\phi}{(1-\phi^2)^{1-\frac{2}{p}}}
\qquad \text{for} \quad p - 2 \in \mathbb{N}\,.
\end{equation}
This can be explicitly integrated to yield 
\begin{equation} \label{xipphi}
\xi_p(\phi) =\, _2F_1\left[\frac{1}{2},1-\frac{2}{p};\frac{3}{2};\phi^2\right]\, \phi\,,
\end{equation}
where $_2F_1(a,b;c;u)$ is the Gauss hypergeometric function (which is well defined on the unit interval for its fourth argument $u$, and whenever $c > a+b$, see e.g. \cite{2007tisp.book.....G}). 
Some special values of $p$ however lead to more nice-looking forms: 
\begin{equation}
\xi_2 = \phi\,, \qquad
\xi_4 =  \arcsin(\phi), \qquad \xi_8 = 2 F\left(\frac{\arcsin(\phi)}{2},\sqrt{2}\right), \qquad
\xi_\infty = \tanh^{-1}\phi
\end{equation}
where $F$ is the elliptic integral of the first kind\footnote{Note that we use here the definition of \cite{2007tisp.book.....G}, i.e. 
$F(\varphi,k) = \int_0^\varphi \frac{d \alpha}{\sqrt{1-k^2 \sin \alpha}}$, which differs from the definition used e.g. in Mathematica~\cite{Mathematica}.}. Note that $\xi_\infty$ just equals the variable $\psi$ defined in Eq.(\ref{phipsimh}). 
The minima $\phi=\pm 1$ of the mexican hat potential are mapped respectively to the following values $\xi_p^{\pm}$ given by 
\begin{equation}
\xi_p^{\pm}\equiv \xi_p\left(\phi = \pm 1\right) =  \pm \frac{\sqrt\pi\, \Gamma\left(\frac{2}{p}\right)}{2\, \Gamma\left(\frac{1}{2}+\frac{2}{p}\right)}
\end{equation}
where we recall that $\Gamma(0) = \infty$ and $\Gamma(1/2) = \sqrt\pi$. As a consequence one sees in particular that for $2<p<\infty$ the minima of the mexican hat potential are sent to finite values of the $\xi_p$ variable. We also have $\xi_p(0)=0$, and one can check that the mapping (\ref{xipphi}) is (monotonic and hence) one to one between $\phi \in [-1,1]$ and $\xi \in[\xi_p^-,\xi_p^+]$. In addition, noticing that $d \xi_b/ d \phi$ diverges in $\phi=\pm 1$, one see that the inverse mapping $\phi = \phi(\xi_p)$ can be naturally extended (for finite $p>2$) to a periodic everywhere smooth, non singular function defined on entire real line and of period $4  \xi^+_p $. In general, this inverse mapping, even though it exists, does not correspond to simple functions, however, this is not true for $p=4$ and $p=8$, for which we have  
\bea
\phi = \sin (\xi_4) \;\;\; {\text{and}} \;\;\; \phi = \sin \left( 2 {\text{am}} (\xi_8/2) \right))= 2 {\text{sn}} (\xi_8/2) {\text cn}(\xi_8/2),
\eea
where am is the so-called {\it amplitude} of the elliptic integral $F$, and sn and cn are the so-called sine-amplitude and cosine-amplitude, and we allow now $\xi_p$ to vary over the entire real line. Obviously the period of the first function above is $2 \pi = 2 \sqrt{\pi} \Gamma(1/2)/ \Gamma(1)$, while the period of the second function is $2 \sqrt{\pi} \Gamma(1/4)/\Gamma(3/4) \sim 10.5$. Of course this is not the only possibility to extend the inverse function beyond the points $\xi_p^{\pm}$, however, choosing this way offers an elegant extension of the family of models (which strictly speaking differ from the ones (\ref{eq:P_our_model}) where the function $(1-\phi^2)^2$ is not periodic). It would be interesting to investigate if this ``periodic'' extension would allow to find  solutions with a non trivial time dependence interpolating between non adjacent minima similarly to what is known to exist in the sine-Gordon model.

\begin{figure}
\centering
\includegraphics[scale=.5]{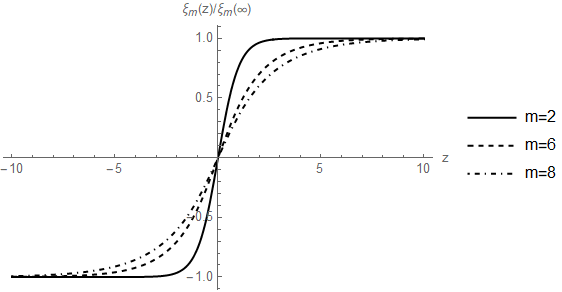}
\caption{Behaviour of the domain wall profile $\xi(z)$ (\ref{eq:xim}) normalized to its value at infinity for different values of $m$.
\label{fig:xi}}
\end{figure}

The interest of the change of variable (\ref{xipphi}) appears considering a Lagrangian of the form (\ref{eq:P_our_model}) and choosing $p = m$. Noting then 
 $\xi = \xi_m$, as well as defining $\Y$ as in (\ref{defX}) replacing there $\phi$ by $\xi$, we get that $P$ now reads
\begin{equation} \label{Lagxinm}
P = 
\frac{\kappa\, (-2\Y)^{m/2}}{2(m-1)}
+ \frac{1-\kappa}{2(n-1)}\left(1-\phi^2\right)^{2\left(1-\frac{n}{m}\right)}\,(-2\Y)^{n/2}
+ \left(1-\phi^2\right)^{2\left(1-\frac{2}{m}\right)}\,\Y\,.
\end{equation}
where $\phi$ is now considered as a function of $\xi$ (i.e. $\phi = \phi(\xi)$ which we can -- but do not have to -- consider as periodic in $\xi$). 
In this form the Lagrangian is no longer singular at the finite values $\phi= \pm 1$ (corresponding to $\xi^{\pm}$), even though the purely "kinetic" term of $\xi$ has the non standard form 
$\propto (-X_\xi)^{m/2}$. For the family (\ref{Lagxinm}), if one notes 
 $w(\xi) = 1-\phi^2(\xi)$, the first integral ${\cal J}$ is found to be (while equations (\ref{eq:J_def})-(\ref{JEeq}) hold, {\it mutatis mutandis})
\begin{equation}
\mathcal{J} =  w^{2\left(1-\frac{2}{m}\right)}\Y
+ \frac{1-\kappa}{2}w^{2\left(1-\frac{n}{m}\right)}(-2\Y)^\frac{n}{2}+ \frac{\kappa}{2}(-2\Y)^\frac{m}{2}\,.
\end{equation}
Explicity, we find the field equation operator ${\cal E}$ given by 
\begin{equation}
\begin{aligned}
\mathcal{E} =& \,
\left[w^{2\left(1-\frac{2}{m}\right)}
- \frac{n(1-\kappa)}{2}\,w^{2\left(1-\frac{n}{m}\right)}\left(\xi'\right)^{n-2}
- \frac{m\kappa}{2}\left(\xi'\right)^{m-2}\right]\xi''\\
& \quad
- 2 \left[\frac{m-2}{m}\,w^{2\left(1-\frac{3}{m}\right)}\left(\xi'\right)^2
- \frac{(m-n)(1-\kappa)}{m}\,w^{2\left(1-\frac{n+1}{m}\right)}\left(\xi'\right)^n\right]\,\phi(\xi)\,,
\end{aligned}
\end{equation}
which in particular, as we have $2 <n<m$, implying $m \geq n+1> 3$, is nowhere singular. 
The domain wall profile, solution of the above, is obviously given by 
\bea\label{eq:xim}
\xi(z) =\,  _2F_1\left[\frac{1}{2},1-\frac{2}{m},\frac{3}{2},\tanh^2(z)\right]\, \tanh(z)\,.
\eea
Those profiles are shown in figure \ref{fig:xi} for the cases $m=2$ (the usual $\tanh$), $m = 6$ and $m = 8$.\\

\subsection{Energy, Bogomolny and topological considerations} \label{ENERBOGO}
Before writing out in the next section the explicit theory of perturbations around our kinks, we would like here to study their energy making the link with Bogomolny's and Derrick's arguments. 
To that hand we first consider the theory written in the $\psi$ variable, and start with the general form (\ref{psigen}) which encompasses the canonical mexican hat model (allowing for a non vanishing $\kappa_0$). The total energy density ${\cal H}(t,z)$  of a given (arbitrary) field configuration is easily found to be 
\begin{equation} \label{Hamiltkappa}
 {\cal H}(t,z) = -\left(\sum_{n \in \mathbb{N}}
\kappa_{2 n} (\psi'^2-\dot{\psi}^2)^{n} + 2 n \kappa_{2 n} \dot{\psi}^2 (\psi'^2-\dot{\psi}^2)^{n-1}
\right)\cosh^{-4} \psi
\end{equation}
where to simplify the discussion we assume here and henceforth that only $\kappa_{k}$ with even $k$ are non zero. 
In the case of the canonical mexican hat model (recall that we just have then $\kappa_0=\kappa_2=-1/2$) we find an energy density ${\cal H}(t,z) = (1+\psi'^2+\dot{\psi}^2)/(2 \cosh^4 \psi)$.
Using then the notation 
\bea
\x &=& \psi' \\
\y &=& \dot{\psi} 
\eea
We see that the Bogomolny trick  and decomposition (\ref{kinBog}-\ref{topBog}) amounts here to just write the polynomial in $\x$ and $\y$ appearing in the numerator of ${\cal H} = (1+\x^2+\y^2)/(2 \cosh^4 \psi)$
as 
\bea \label{polydecmh}
1+\x^2+\y^2 = \y^2 + (\x\pm 1)^2 \mp 2\x, 
\eea
 where the first term on the right hand side yields the kinetic energy (after the proper division by $2 \cosh^4 \psi$), the second one vanishes for the wall profile $\x=\mp 1$ and the last one give the equivalent of the "topological" charge (\ref{topBog}), i.e. it gives choosing here the lowest sign (as would be appropriate for the kink, as opposed to the antikink which would correspond to the solution $x=-1$ and the choice of the upper signs)
\bea \label{topchargedetails}
\int_{-\infty}^{+\infty} \frac{ \psi'} {\cosh^4 \psi} dz &=&  \int_{\psi(-\infty)}^{\psi(+\infty)} \frac{ d u } {\cosh^4 u} \\
&=&  \left[\tanh u - \frac{1}{3} \tanh^3 u\right]_{u= \psi(-\infty)}^{u=\psi(+\infty)}\\
&=& \left[u - \frac{1}{3} u\right]_{u=\phi(-\infty)}^{u=\phi(+\infty)} \\
&=& \int_{\phi(-\infty)}^{\phi(+\infty)}(1-u^2) du
\eea
where the last form indeed matches the expression (\ref{topBog}) with the mexican hat potential (\ref{mhpot}). For the domain wall profile we find the the above expression yield 4/3 (see eq. (\ref{intek})). We now show that a decomposition similar to (\ref{polydecmh}) exists in general for our theories. Indeed, considering (\ref{Hamiltkappa}) we see that the polynomial equivalent to (\ref{polydecmh}) reads in full generality
\bea
\Pi_0(\x,\y) = - 2 \left( \sum_{n \in \mathbb{N}}
\kappa_{2 n} (\x^2-\y^2)^{n} + 2 n \kappa_{2 n} \y^2 (\x^2-\y^2)^{n-1} \right)
\eea
so that the Hamiltonian density is just $\Pi_0(x,y)/ 2 \cosh^4 \psi $, while, in order to have a domain wall with profile $\psi = \pm z$, the coefficient $\kappa_n$ must obey (see equation (\ref{constraintkappa})) 
\bea \label{constraintkappafinal}
\Sigma_{\kappa,0} = 2 \Sigma_{\kappa,1} 
\eea
where the $\Sigma_{\kappa,k}$ are defined by  
\bea
\Sigma_{\kappa,k} = \sum_{n \in \mathbb{N}} \kappa_{2n} n^k,
\eea 
where we imply in particular that $\Sigma_{\kappa,0}  = \sum_{n \in \mathbb{N}} \kappa_{2n}$ (using the convention that $0^0 = 1$). 
At this stage, considering the form of $\Pi_0$, we can notice that the Hamiltonian cannot be bounded below if the largest integer $n$ for which $\kappa_{2n}$ does not vanish, call it $n_{max}$, is even. In constrast, if $n_{max}$ is odd we see that at large $\x$ and $\y$ the dominant terms in $\Pi_0(\x,\y)$ read $(-2 \kappa_{2n_{max}})(x^2 + (2n_{max}-1)y^2)(x^2-y^2)^{n_{max}-1})$ which shows that the Hamiltonian is bounded below for negative $\kappa_{2 n_{max}}$ (and finite $\psi$).  In fact it can further be shown (see below) that it is possible to find, for specific odd $n_{max}$ and $\kappa_{2n}$, an everywhere positive Hamiltonian (the Hamitonian vanishing only at $(x=0,y=0)$). Let us now expand $\Pi_0$ around $x=\pm 1$ and $y=0$ corresponding to the domain wall solution. We find after some simple manipulations 
\bea \label{PolyExpand}
\Pi_0(\x,\y) =\left(4 \Sigma_{\kappa,1}-2\Sigma_{\kappa,0} \right) \mp  4 \x \Sigma_{\kappa,1}  + \Pi\left((\x\mp 1),y^2\right), 
\eea
where $\Pi(a,b)$ is a polynomial in $a$ and $b$ which vanishes in $(a=0,b=0)$ and in addition start only at order $a^2$ and $b$ expanding around this point. We see that the first term on the right hand side of (\ref{PolyExpand}) vanishes by virtue of (\ref{constraintkappafinal}). 
Hence we can write the total energy of any field configuration in a theory of the family (\ref{psigen}) which has a domain wall solution as 
\bea
{\cal H}(t) =  {\cal H}_{kin,grad}(t) + {\cal H}_{\infty}(t), \label{Bogopsi}
\eea
where the two contributions on the right hand side read,  using (\ref{constraintkappafinal}) 
\bea
{\cal H}_{kin,grad}(t) &=& \int \frac{\Pi\left((\psi'\mp1),\dot{\psi}^2\right)}{2 \cosh^4 \psi} dz \\
{\cal H}_{\infty}(t) &=&  \mp \Sigma_{\kappa,0}  \int \frac{\psi'}{\cosh^4 \psi} dz \\
\eea
where one sees that the last term is a topological conserved charge just identical (up to a constant factor) to the one of the canonical model (\ref{topchargedetails}) (see also (\ref{topBog})).
In the variable $\psi$ it is associated with the current 
\bea
\tilde{J}^{\mu}_\psi = \tilde{\normconst} \epsilon^{\mu \nu} \partial_{\nu}\left( \int_{\psi_0}^\psi \frac{du}{\cosh^4 u}\right).
\eea
The above decomposition (\ref{Bogopsi}) generalizes the one of Bogomolny in our context, and one can check that with the choice of non vanishing $\kappa_{2n}$ given by $\kappa_0=\kappa_2=-1/2$ we find back exactly the form (\ref{Bogolboundgen}). It also allows to check for the stability of the wall configuration within a class of field configuration sharing the same conserved charge ${\cal H}_{\infty}$. To that end we can look at the the behaviour of the the contribution ${\cal H}_{kin,grad}(t)$ by expanding $\Pi$ around $(0,0)$. Specifically, taking into account the constraint (\ref{constraintkappafinal}) we find the following expansion of $\Pi_0$: 
\bea \label{expandPIzero}
\Pi_0(\x,\y) = \mp 2 \x \Sigma_{\kappa,0}  - (x\mp 1)^2 \left(4 \Sigma_{\kappa,2} - \Sigma_{\kappa,0}\right)
- y^2 \Sigma_{\kappa,0}  + ... 
\eea
where the left over terms are at least cubic in $(x\mp 1)$ and $y$. This shows that the domain wall solution represent a local minimum of the energy 
in the class of all field configuration having the same topological charge 
provided that the quantities $\Sigma_{\kappa,0}$ and $\Sigma_{\kappa,2}$ (defined above)  verify 
\bea
4 \Sigma_ {\kappa,2} < \Sigma_{\kappa,0}&< & 0 \label{conscons}
\eea
For the domain wall one has $\Pi=0$ (i.e. $\Pi(\pm 1, 0) =0$) which means that the energy is only containing a non zero topological contribution ${\cal H}_{\infty}$.
Note however that in contrast to the canonical mexican hat domain wall, the domain wall "without a potential" (i.e. whenever $\kappa_0$ vanishes) can not be global minimum of the energy within the class of configuration with the same topological charge. Indeed, from the above discussion we ses that 
$\Pi = \Pi_0 \pm 2  x \Sigma_{\kappa,0}$, but $\Pi_0$ vanishes in $(x=0,y=0)$ where the dominant terms as $\x$ and $\y$ approach zero are quadratic in $\x$ and $\y$. This means that $\Pi$ has to change sign accross $(x=0,y=0)$ and must be somewhere negative, preventing the local minimum of $\Pi$ at $x=\pm1$, $y=0$ (where $\Pi$ vanishes) to be a global minimum.

\begin{figure}
\centering
\includegraphics[scale=.5]{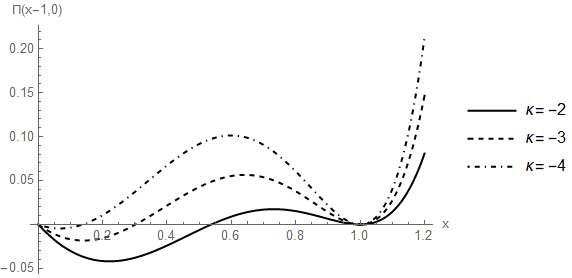}
\caption{Behaviour of $\Pi(x-1,y^2)$ on the $y = 0$  line, for the model (\ref{Pnmpsi}) with $(n,m) = (4,6)$ and different values of $\kappa$.
\label{fig:Pi}}
\end{figure}

\begin{figure}
\centering
\includegraphics[scale=.5]{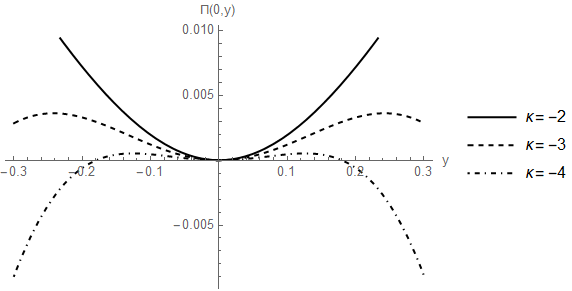}
\caption{Behaviour of $\Pi(x-1,y^2)$ on the $x = 1$  line, for the model (\ref{Pnmpsi}) with $(n,m) = (4,6)$ and different values of $\kappa$.
\label{fig:Pi_y}}
\end{figure}

Conditions (\ref{conscons}) are the ones $\kappa_n$ should obey in order to get a stable wall configuration.
Setting $\kappa_2$, $\kappa_{n}$ and $\kappa_m$ as in equations (\ref{kappa2}), (\ref{kappan}) and (\ref{kappam}) for some specific even $n$ and $m$, we can check that above conditions (\ref{conscons}) are equivalent to conditions (\ref{eq:P_our_modelbound}) for the set of models  (\ref{Pnmpsi}).
To discuss a more explicit case, let us consider the simple model in the class (\ref{eq:P_our_model}) with there $n=4$ and $m=6$. Explicitly the Lagrangian of this model reads in the $\psi $ variable 
\bea \label{Lagnm46psi}
\frac{1}{\cosh^4\psi}\left(X_\psi + \frac{1-\kappa}{6}\left(-2X_\psi\right)^{2}+ \frac{\kappa}{10}\left(-2X_\psi\right)^{3}\right)\,,
\eea
hence we have $\kappa_2 = -1/2$, $\kappa_4 = (1-\kappa)/6$, $\kappa_6 = \kappa/10$, so that $\Sigma_{\kappa,0} = -(5+\kappa)/15$ and $4 \Sigma_{\kappa,2}-\Sigma_{\kappa,0}=1+\kappa$, hence the constraint 
(\ref{conscons}) is satisfied provided that 
\bea \label{boundkappa51}
 -5< \kappa<-1.
\eea This range corresponds also to the allowed range for $\kappa$ given in Eq.(\ref{eq:P_our_modelbound}). Moreover, in the line of the discussion following equation (\ref{constraintkappafinal})
one can show that restricting further $\kappa$ to be larger than $-(17+3 \sqrt{21})/10 \sim -3.07$ we get an everywhere positive Hamiltonian ${\cal H}(t,z)$.
As further expected, we find in that case that $\Pi$ vanishes at $x=\pm 1, y=0$ which is a local minimum of $\Pi$, but $\Pi$ is negative somewhere on the $y=0$ line in the $(x,y)$ plane and hence 
$x=\pm 1, y=0$ is not a global minimum of $\Pi$. This is shown in figure \ref{fig:Pi} for different values of $\kappa$, while figure \ref{fig:Pi_y} shows the shape of the polynominal $\Pi$ along the $x = 1$ line.

It is also interesting to see how the usual scaling argument due to Derrick \cite{Derrick:1964ww} applies here. To that end, consider a rescaling of the domain wall solution $\psi_0(z) = \pm z$ as in
$\psi_{\omega} =\psi_0(\omega z)$. The total energy of the rescaled field configuration $\psi_\omega$ is easily obtained as 
\begin{equation}
\mathcal{H}_\omega = - \frac{4}{3}\,\sum_{n\in\mathbb{N}}\omega^{n-1}\kappa_n\,.
\end{equation}
Restricting ourselves here to the case of even $n$ we obtain easily the first and second derivatives of $\mathcal{H}_\omega$ evaluated at $\omega = 1$ as
\begin{equation}
\left.\frac{d\mathcal{H}_\omega}{d\omega}\right\vert_{\omega = 1} = - \frac{4}{3}\left(2\Sigma_{\kappa,1}-\Sigma_{\kappa,0}\right)
\qquad
\left.\frac{d^2\mathcal{H}_\omega}{d\omega^2}\right\vert_{\omega = 1} = - \frac{4}{3}\left(4\Sigma_{\kappa,2}-6\Sigma_{\kappa,1}+3\Sigma_{\kappa,0}\right)\,.
\end{equation}
The first derivative above vanishes by virtue of the relation (\ref{constraintkappafinal}), thus confirming that the domain wall is indeed a solution.
The second derivative is positive if the condition (\ref{conscons}) holds, thus Derrick's usual scaling no-go argument is evaded and the domain wall is stable against dilatations.

\subsection{Static and moving walls}
In the canonical mexican hat model (\ref{eq:P_can_genV})-(\ref{mhpot}), equation (\ref{constraintkappa}) has only the roots $\lambda = \pm 1$. However, considering the more general models  
 (\ref{eq:P_our_model}) (or (\ref{Pnmpsi})), it is possible that equation (\ref{constraintkappa}), which now reads  
\bea \label{lambdaeqbis} \label{eq:lambda_def}
\lambda^2+(\kappa-1)|\lambda|^n-\kappa\,|\lambda|^m=0,
\eea
has some other roots $\lambda$ different from $\pm 1$. This would yield a domain wall solution of profile 
\bea \label{lambdaprofile}
\phi_\lambda(z) = \pm \tanh\left(\lambda\,z \right).
\eea
Note however, that $\lambda =\pm 1$ is always a solution of equation (\ref{lambdaeqbis}), so that the standard domain wall profile coexists always with the profile (\ref{lambdaprofile}). E.g., the Lagrangian 
(\ref{Lagnm46psi}) admits, beyond the "canonical" wall $\phi=\tanh(\pm z)$ another wall solution of the kind (\ref{lambdaprofile}) with $\lambda = \pm 1/\sqrt{-\kappa}$.
However, while properties of the solution (\ref{lambdaprofile}) (with $\lambda \neq \pm 1$) are given in appendix \cref{sec:app_lambda}, it is also shown there that 
both solutions cannot be stable simultaneously: the solution (\ref{lambdaprofile}) can be made stable at the price of violating the bounds  \eqref{eq:P_our_modelbound} on $\kappa$ which are in turn necessary for the stability of the solution with the canonical profile. However, having more than three terms in the Lagrangian 
(\ref{eq:P_our_model}) (or (\ref{Pnmpsi})) leads to the possibility to have more roots to the equation (\ref{constraintkappa}) and hence possibly more than one stable wall solution, this will be investigated elsewhere. 

Another possibility to extend the solutions discussed above is to let the walls move. In particular, using the $\psi$ variable and considering for simplicity the models (\ref{Pnmpsi}), it is easy to see that the part of the field equations that do not contain any second derivatives, is in full generality proportional (and as consequence of Lorentz invariance) to 
\bea
X_{\psi} + \frac{1-\kappa}{2}\left(-2 X_{\psi}\right)^{n/2} + \frac{\kappa}{2}\left(-2 X_\psi\right)^{m/2}
\eea
which for a static wall is in turn proportional to the expression of the first integral ${\cal J}$. This means that any static wall profile (\ref{lambdaprofile}) extends (including the "canonical" case $\lambda = \pm 1 $) to a moving solution of the form 
\begin{equation}
\phi_m(t,x) = \pm \tanh  \left(\lambda \frac{z \pm \beta t}{\sqrt{1-\beta^2}} \right)\,,
\end{equation}
where $\beta <1$ is the dimensionless speed, and where one has $ - 2 X_{\psi} = \psi'^2 - \dot{\psi}^2 \approx \lambda^2$.

\subsection{Sine-Gordon like and other walls}
\label{sec:sGdomainwall}

The above discussion and construction can easily be extended to other kind of kink profiles such as the one of sine-Gordon or more generally the family of models (\ref{LKk}). Indeed, consider Lagrangians of the form 
\bea \label{psigenmoregeneral}
{\cal L} = \left(\sum_{n \in \mathbb{N}}
\kappa_n (-2 X_\psi)^{n/2}\right)\cosh^{-2k} \psi.
\eea
As the discussion of sections \ref{sec:changing_var} and \ref{ENERBOGO} applies whatever the $\psi$-dependent factor in front of the Lagrangian, its conclusions hold also for the family (\ref{psigenmoregeneral}). In particular, $\psi =\lambda z $ is a solution as long as $\lambda$ obeys (\ref{constraintkappa}) and moreover, the solution with $\lambda = \pm 1$ is stable provided conditions (\ref{constraintkappafinal}) and (\ref{conscons}) hold. Turning back to the original $\phi$ variable, the corresponding Lagrangians are simply given by (\ref{moregeneralform}), where the powers of $(1-\phi^2)$ are replaced by powers of $\vert \phi'\vert$, considered and expressed in terms of $\phi$, and it is easy to get the corresponding domain wall profiles for the $\phi$ variable. In particular, for $k=1$, we get stable domain wall profiles identical to the one of sine-Gordon model reading as in eq. (\ref{sGprofile}) with the Lagrangians
\bea \label{moregeneralformsG}
P = \sum_{n\geq 2}\kappa_{2n} \sin(\phi/2-p\pi)^{2-2n} (-2X)^n
\qquad p \in \mathbb{Z}\,.
\eea

One feature of the sine-Gordon model is its integrability leading in particular to non trivial solutions such as breathers or kink-antikink (see e.g. \cite{Shnir:2018yzp}). It would be interesting to investigate if some remnant of such solutions still exist in the kind of models considered here.

\section{Wall perturbations}
\label{pertsection}
We focus here on properties of perturbations around the domain wall solutions discuss in the previous section. To be specific, we will concentrate on the set of theories (\ref{eq:P_our_model})-(\ref{Pnmpsi})-(\ref{Lagxinm}).

\subsection{Quadratic perturbations}
We write a generic field configuration as $\phi(t,z) = \phi(z) + \varphi(t,z)$, where $\phi(z) = \pm \tanh z$  and the domain wall perturbations $\varphi(t,z)$, once Fourier transformed w.r.t. time, obey equation 
(\ref{eq:SL_pert}). We give in the table \ref{tbl:lin_pert} below the relevant coefficients $\mathcal{Z}^{00}$, $\mathcal{Z}^{zz}$ and $\mathcal{M}^2$ appearing in this equation, we also indicated there the value of the energy density ${\cal H}(z)$ of the domain wall solution. These functions are given both for the generic $P_{n,m}$ Lagrangians of equation (\ref{eq:P_our_model}), for the specific choice 
$(n,m) = (4,6)$ corresponding to the Lagrangian $P_{4,6}$, and for the canonical mexican hat model  (\ref{eq:P_can_genV})-(\ref{mhpot})
whose Lagrangian is denoted by $P_\text{can}$. The quantities relevant for this last model are henceforth indicated with an index "$_{\text{can}}$".
\renewcommand{\arraystretch}{2}
\begin{table*}[h]
	\begin{center}
		\begin{tabular}{| c || c | c | c |}
			\hline
			 & $P_{n,m}$ &$P_{4,6}$  & $P_\text{can}$ \\
			\hline \hline
			$\mathcal{H}(z)$ & $\frac{1}{2}\left(\frac{n-2}{n-1}+\frac{(m-n)\kappa}{(n-1)(m-1)}\right)\cosh^{-4}(z)$ & $\frac{5+\kappa}{15}\, \cosh^{-4}(z)$ & $\cosh^{-4}(z)$ \\ 
			\hline\hline
			$\mathcal{Z}^{00}(z)$ & $-\frac{1}{2}\left(\frac{n-2}{n-1}+\frac{(m-n)\kappa}{(n-1)(m-1)}\right)$ & $-\frac{5+\kappa}{15}$ & $-1$  \\ 
			\hline
			$\mathcal{Z}^{zz}(z)$ & $\frac{2-n+(n-m)\kappa}{2}$ & $ -\left(1+\kappa\right)$ & 1  \\ 
			\hline
			$\mathcal{M}^2$(z)  & $\left(2-n+(n-m)\kappa\right)\left(3\phi^2(z)-1\right)$ & $-2\left(1+\kappa\right)\left(3\phi^2(z)-1\right)$ & $2\left(3\phi^2(z)-1\right)$ \\ 
			\hline 
		\end{tabular}
		\caption{Comparison of the energy density and the perturbative quantities for the generic $P_{n,m}$ Lagrangians of equation (\ref{eq:P_our_model}), for the specific choice 
$(n,m) = (4,6)$ corresponding to the Lagrangian $P_{4,6}$, and for the canonical mexican hat model. \label{tbl:lin_pert}}
	\end{center}
\end{table*}
We first note that the perturbations of our "domain walls without a potential" discussed here have an action very similar to the ones of the canonical mexican hat wall. In particular, the kinetic matrix is constant, thus the perturbations are well defined everywhere in space. More precisely, we see that, as $\mathcal{M}^2 = \mathcal{Z}^{zz}\,\mathcal{M}^2_\text{can}$ and $\mathcal{Z}^{zz}$ is constant, a perturbation of our wall obeys the same equation as a perturbation of the canonical wall
\begin{equation}
\left(\mathcal{Z}^{zz}_\text{can}\,\varphi_k'\right)' 
-\left(\mathcal{Z}^{00}_\text{can} \,\tilde{\omega}_k^2+\mathcal{M}^2_\text{can}\right)\varphi_k = 0\,,
\end{equation}
where the frequency of each mode is multiplied by an universal factor obtained below
\begin{equation}
\tilde{\omega}_k^2 = \frac{(n-2)(m-1)+(m-n)\kappa}{\left(2-n+(n-m)\kappa\right)(n-1)(m-1)} \; \omega_k^2 \,.
\end{equation}
In order to find stable perturbations, we recall that we have to demand that conditions (\ref{eq:conds}) are obeyed, which amounts to just demand that $\mathcal{Z}^{00}$ is negative and $\mathcal{Z}^{zz}$ positive. In turn, this gives the bounds on $\kappa$ given in equation (\ref{eq:P_our_modelbound}).

We can further note that our models allows to find walls which have exactly the same profile and energy density as the canonical walls by tuning to 1 the coefficient in front of $\cosh^{-4}(z)$ in ${\cal H}(z)$ choosing $\kappa=n(m-1)/(m-n)$. These walls are thus perfect "Doppelg\"anger" walls to use the terminology of \cite{Andrews:2010eh}. However, for such walls, the bounds (\ref{eq:P_our_modelbound}) are violated so in our cases these perfect Doppelg\"anger walls are not stable. However, choosing 
\bea
\kappa  = \frac{n(m-1)(n-2)}{(n-m)(2-n+m(n-1))},
\eea
which satisfies the bounds (\ref{eq:P_our_modelbound}), we get $\tilde{\omega}_k^2 = \omega_k^2$ and so the theory has exactly the same spectrum as the canonical one. This correspond explicitly to the family of Lagrangians 
\begin{equation} \label{mimickers}
P(\phi,X) = X + \frac{1}{2(m-n)(2-n+m(n-1)}\left[\frac{m(m-2)\left(-2X\right)^{n/2}}{\left(1-\phi^2\right)^{n-2}}
-\frac{n(n-2)\left(-2X\right)^{m/2}}{\left(1-\phi^2\right)^{m-2}}\right]\,,
\end{equation}
which have stable domain walls with profile identical to the one of the canonical mexican hat, and an energy density, and kinetic matrix just rescaled by a common factor given by 
\bea
\frac{(n-2)(m-2)}{2(2-n+m(n-1))}=1-\frac{m n}{2( m n -(m+n) +2)}.
\eea
In this class of models, that we will call here and henceforth a mimicker, the simplest ones are possibly obtained by choosing $(n,m) = (4,6)$ and $\kappa=-5/4$ yielding the simple Lagrangian 
\begin{equation}\label{eq:case_n4_m6}
P(\phi,X) = X + \frac{3\,X^2}{2\left(1-\phi^2\right)^2} + \frac{X^3}{\left(1-\phi^2\right)^4}\,,
\end{equation}
which has a domain wall solution $\phi = \pm \tanh(z)$, a Hamiltonian everywhere positive as seen in the previous section, and energy density and kinetic matrix just rescaled by a global factor $1/4$ with respect to the canonical ones.
Note that for this particular model, we can compute 
\bea
P_X (1-\phi^2)^4 &=& \left((1-\phi^2)^2+\frac{3}{2} X\right)^2 + \frac{3}{4} X^2\\
(2 X P_{XX}+ P_X) (1-\phi^2)^4 &=& \left((1-\phi^2)^2+\frac{9}{2} X\right)^2 - \frac{21}{4} X^2
\eea
so that we see that condition (\ref{cond111}) is always fullfilled in agreement with having an everywhere positive Hamiltonian, while condition (\ref{cond222}) can be violated somewhere in the field space. However, the later condition is verified on the wall background and in its vicinity in agreement with the found local stability.

\subsection{Cubic perturbations and strong coupling}
As we saw in the previous section, the walls considered here are local minima of the energy in the class of field configuration with fixed boundary conditions at $z=\pm \infty$. This contrasts with canonical domain walls which are global minima. As a consequence, one should be able to distinguish the two looking at higher order perturbations as we now show.  Up to surface terms,
for a generic theory of the kind (\ref{defLagrangian}), the third-order perturbed Lagrangian reads
\begin{equation}
\delta^{(3)}\mathcal{L} = - \frac{1}{3!}\left[ \mathcal{Y}^{\mu\nu\rho}\,\partial_\mu\varphi\,\partial_\nu\varphi\,\partial_\rho\varphi - 3\mathcal{Y}^{\mu\nu} \varphi\,\partial_\mu\varphi\,\partial_\nu\varphi + \mathcal{Y}\,\varphi^3\right]\,,
\end{equation}
where the different coefficients appearing above are given by 
\begin{subequations}
\begin{align}
& \mathcal{Y}^{\mu\nu\rho} =
P_{XXX}\,\partial^\mu\phi\,\partial^\nu\phi\,\partial^\rho\phi-3P_{XX}\,\eta^{\mu\nu}\,\partial^\rho\phi\,,\\
& \mathcal{Y}^{\mu\nu} =
P_{XX\phi}\,\partial^\mu\phi\,\partial^\nu\phi-P_{X\phi}\,\eta^{\mu\nu}\,,\\
& \mathcal{Y} =
-P_{\phi\phi\phi}-\partial_\mu\left(P_{X\phi\phi}\,\partial^\mu\phi\right)\,.
\end{align}
\end{subequations}
For the canonical model (\ref{eq:P_can_genV})-(\ref{mhpot}), only $\mathcal{Y} = V_{\phi\phi\phi} = 12\phi$ is non-vanishing. For the $P_{n,m}$ models (\ref{eq:P_our_model}), as well as their 
subset mimickers (\ref{mimickers}), one finds for the background given by the wall of canonical profile
$\phi = \pm \tanh(z)$, in particular some relevant coefficients are gathered in the following table \ref{tbl:cubiccoef}.
\begin{table*}[h]
	\begin{center}
		\begin{tabular}{| c || c | c | c |}
			\hline
			 & Generic $P_{n,m}$  & Mimicker $P_{n,m}$ & Mimicker $P_{4,6}$  \\
			\hline \hline
			$\mathcal{Y}^{00z}$ & $-\frac{3}{2}\left(\frac{n(n-2)(\kappa-1)}{n-1}-\frac{m(m-2)\kappa}{m-1}\right)\frac{1}{1-\phi^2}$ & 0 & 0 \\ 
			\hline
			$\mathcal{Y}^{zzz}$ & $\frac{n(n-2)(\kappa-1)-m(m-2)\kappa}{2} \frac{1}{1-\phi^2}$ & $\frac{mn(n-2)(m-2)}{2(2-n+m(n-1))}\frac{1}{1-\phi^2} $ & $\frac{6}{1-\phi^2}$ \\ 
			\hline\hline
			$\mathcal{Y}^{00}$ & $\left(\frac{n(n-2)(\kappa-1)}{n-1}-\frac{m(m-2)\kappa}{m-1}\right)\frac{\phi}{1-\phi^2}$ & 0 & 0 \\ 
			\hline
			$\mathcal{Y}^{zz}$ & $-\left(n(n-2)(\kappa-1)-m(m-2)\kappa\right)\frac{\phi}{1-\phi^2}$ & $-\frac{mn(n-2)(m-2)}{(2-n+m(n-1))}\frac{\phi}{1-\phi^2}$ & $-\frac{12 \phi}{1-\phi^2}$\\
			\hline
		\end{tabular}
		\caption{Some relevant coefficients of the cubic vertices for the generic $P_{n,m}$ Lagrangians of equation (\ref{eq:P_our_model}), for the subset of mimicker models, and for the specific choice 
$(n,m) = (4,6)$. \label{tbl:cubiccoef}}
	\end{center}
\end{table*}
One can notice that for all $P_{n,m}$ models, $\mathcal{Y}^{00} = -2\phi/3\, \mathcal{Y}^{00z}$ and $\mathcal{Y}^{zz} = -2\phi\, \mathcal{Y}^{zzz}$. Moreover, for the mimickers, all contributions containing time derivatives of the perturbations vanish at cubic order. However, the cubic interactions are found diverging at large $z$, for which, for the domain wall profile, $1/(1-\phi^2)$ as well as 
$\phi/(1-\phi^2)$ diverge. Hence the perturbation theory in the $\phi$ variable diverges at large $z$ off the wall. 
Note however, that as we have shown that the wall is a local minimum of the energy in the class of field configurations with fixed boundary conditions, one expects that there is a range of localized perturbations of the wall which are absolutely stable. To end, we also notice that one cannot mimic our models with a $P(\phi,X)$ of the form  $f(X) - V(\phi)$ as $\mathcal{Y}^{\mu\nu}$ would be vanishing.
Note also that the generic properties of the perturbations found in this section using the $\phi$ variable: sound quadratic perturbations, off-the-wall strong coupling at cubic order, persists e.g. if one trade the $\phi$ variable to $\xi$ (once the quadratic perturbations properly normalized).

\section{Conclusion}
\label{sec:concl}

In this work we have studied domain walls in some k-essence theories. We have shown in particular that domain walls can be supported by non-canonical kinetic terms only, without the help of a potential.
If pure $P(X)$ theories cannot accommodate these unidimensional solitons, the class of Lagrangians \eqref{eq:P_our_model} is an example of potential-free theories that can, and we have obtained an even  larger set of theories sharing the same property. Moreover, we showed that theories can be found having domain wall profiles just identical to the ones of canonical field theories such as a canonical scalar field with a mexican hat potential or sine-Gordon theories. We have also showed that our walls are local minima of the energy in the set of field configurations with some fixed topological charge, however, in contrast with the usual case, they are not global minima. We also studied the quadratic perturbations of these walls, showing in particular that these perturbations can be stable and even identical to the perturbations of the domain walls of canonical models. Canonical walls can however be distinguished from the one discovered here looking at cubic vertices of the the perturbations, which in our case become strong off the wall surface. 

This work raises various questions beyond the ones already mentioned in the main text above. First, as it is clear that our walls are only stable when subjected to small enough and localized perturbations (hence "perturbatively stable"), it would be interesting to study their classical or quantum decay. One could also imagine constructing similar objects in a more general setup such as Horndeski theories or studying the possibility to get solitons with different topologies (such as strings or monopoles) and higher dimensions along the line considered here. On a more phenomenological account, it is known that k-essence can have interesting application in the early Universe, e.g. during inflation (see e.g. 
\cite{ArmendarizPicon:1999rj,ArmendarizPicon:2000yi,Garriga:1999vw,kinfPlanck,kinfTsujikawa,Martin_2013}), an interesting question would hence to look there at the possibility of the formation and decay of the kind of domain walls considered here in the early times, and a related question would be to study the effects of turning on gravity. 

\acknowledgments 

The authors would like to thank Eugeny Babichev, Thibault Damour, Gilles Esposito-Far\`{e}se, J\'er\^ome Martin, Slava Mukhanov and Daniele Steer for very enlightening discussions.

\appendix



\section{DBI-inspired model}
\label{sec:app_DBI}

In this appendix, we construct another potential-free theory admitting stable hyperbolic tangent solutions, inspired by the DBI Lagrangian $c(\phi)\sqrt{1+2X/c(\phi)}$. Let's consider the theory
\begin{equation}
P(\phi,X) = P_0 \sqrt{1+\frac{2X}{c(\phi)} +\alpha_n(\phi)\,\left(-2X\right)^{n/2}}\,,
\end{equation}
with a constant $P_0$ and an integer $n > 2$. Then
\begin{equation}
\mathcal{J} = -P_0 \left(1-\frac{n-2}{2}\,\alpha_n\,\left(-2X\right)^{n/2} \right)\left[1+\frac{2X}{c} + \alpha_n\,\left(-2X\right)^{n/2}\right]^{-1/2} \approx 0
\end{equation}
is solved by $\alpha_n(\phi) = \frac{2}{n-2}\left(1-\phi^2\right)^{-n}$ and the energy density reads
\begin{equation}
\mathcal{H}(z) = -P_0\sqrt{\frac{n}{n-2} -\frac{(1-\phi^2)^2}{c(\phi)}}\,,
\end{equation}
which is defined as long as $(1-\phi^2)^2/c(\phi) < n/(n-2)$. The coefficients of the kinetic matrix are given by
\begin{equation}
\mathcal{Z}^{00} = \frac{P_0}{(1-\phi^2)^2}\sqrt{\frac{n}{n-2} -\frac{(1-\phi^2)^2}{c(\phi)}}
\qquad
\mathcal{Z}^{00} = -\frac{nP_0}{(1-\phi^2)^2}\left(\frac{n}{n-2} -\frac{(1-\phi^2)^2}{c(\phi)}\right)^{-1/2}\,,
\end{equation}
so that the condition (\ref{mathcalG}) is automatically satisfied and the condition (\ref{mathcalN}) is again equivalent to the finiteness of the total energy.
In order to fulfill it, it is easy to see that $P_0$ has to be negative, and that we have to choose carefully $c(\phi)$. For example, the Lagrangian
\begin{equation}
P = P_0\sqrt{1+ \frac{2n}{n-2}\frac{\phi^2(2-\phi^2)}{(1-\phi^2)^2}\,X + \frac{2}{n-2}\frac{(-2X)^{n/2}}{(1-\phi^2)^n}}
\end{equation}
with $P_0$ a strictly negative constant, and $n $ an integer strictly greater than 1, admits stable domain wall configurations.

\section{$\lambda \neq 1$ branch}
\label{sec:app_lambda}

In this appendix, we investigate the $\lambda \neq 1$ branch that was discovered in \cref{eq:lambda_def}. 
Let's recall that, in addition to the usual $\phi_c = \pm \tanh z$ solution, the model \eqref{eq:P_our_model} also accommodates different solutions, given by
\begin{equation}
\phi_\lambda(z) = \pm\tanh\left(\lambda\,z \right)\,,
\qquad \text{where} \qquad
\lambda^2+(\kappa-1)\vert\lambda\vert^n-\kappa\,\vert\lambda\vert^m=0\,.
\end{equation}
For example, in the simplest $(n,m)=(4,6)$ case, this other solutions is given by $\lambda_{4,6} = 1/\sqrt{-\kappa}$.

When $\lambda \neq 1$, we can express $\kappa$ in terms of $\lambda$ as $\kappa = (\vert\lambda\vert^n-\lambda^2)/(\vert\lambda\vert^n-\vert\lambda\vert^m)$\emph{ie.} viewing this condition as a tuning on the Lagrangian to accommodate a given $\lambda$.
In this view, the configurations $\lambda = 1$ and $\lambda \neq 1$ are simultaneously stable iff the $\kappa$ that accommodates the stable $\lambda \neq 1$ solution lies within the bounds (\ref{eq:P_our_modelbound}).

The energy, kinetic matrix and effective mass are then given by
\begin{equation}
\mathcal{H} = \mathcal{A}\,\mathcal{H}_\text{can}\,,
\qquad \mathcal{Z}^{00} = -\lambda^{-2}\mathcal{A}\,,
\qquad \mathcal{Z}^{zz} = \lambda^{-2}\mathcal{B}
\qquad \text{and} \qquad
\mathcal{M}^2 = \mathcal{B}\,\mathcal{M}^2_\text{can}\,,
\end{equation}
where
\begin{subequations}
\begin{align}
& \mathcal{A} = \frac{1}{2} \left(\frac{m-2}{m-1}\,\vert\lambda\vert^{m+2}-\frac{n-2}{n-1}\,\vert\lambda\vert^{n+2}+\frac{n-m}{(n-1)(m-1)}\,\vert\lambda\vert^{n+m}\right)\,\frac{1}{\vert\lambda\vert^m-\vert\lambda\vert^n}\,,\\
& \mathcal{B} =-\frac{1}{2} \,\frac{(m-2)\,\vert\lambda\vert^{m+2}-(n-2)\,\vert\lambda\vert^{n+2}+(n-m)\,\vert\lambda\vert^{n+m}}{\vert\lambda\vert^m-\vert\lambda\vert^n}\,.
\end{align}
\end{subequations}
Let's note that, as in the $\lambda = 1$ case, the spectrum of the perturbations is simply shifted \emph{wrt.} the canonical one
\begin{equation}
\tilde{\omega}_k^2 = \frac{\mathcal{A}}{\mathcal{B}}\, \omega_k^2\,.
\end{equation}
In order to have a stable configuration, we have to impose that $\mathcal{A}$ and $\mathcal{B}$ are simultaneously positive. $\mathcal{B}$ is positive for $\vert\lambda\vert > 1$, whatever the values of $n$ and $m$. $\mathcal{A}$ stays positive for $\vert\lambda\vert$ lying within 0 and some value, say $\bar\lambda$, greater than 1 (indeed $\mathcal{A}(\lambda = 1) = \frac{(n-2)(m-2)}{2(n-1)(m-1)}>0$). Thus there exists always a range $]1,\bar\lambda[$ in which the configuration $\phi_\lambda$ is stable.

However the two configurations with $\lambda = 1$ and $\lambda \neq 1$ cannot be simultaneously stable. In fact $\frac{n-m}{n-2}\kappa - 1$ stays negative for $\vert\lambda\vert>1$,\footnote{Notably around $\lambda = 1$ it comes $\frac{n-m}{n-2}\kappa \simeq 1 - \frac{m-2}{2}\left(\lambda-1\right) + \ldots$} and thus the lower bound of the condition \eqref{eq:P_our_modelbound} when the $\lambda \neq 1$ configuration is stable.
For example in the $(n,m) = (4,6)$ case, the $\lambda = 1$ configuration is stable for $-5<\kappa<-1$ and the $\lambda = 1/\sqrt{-\kappa}$ one, is stable for $-1<\kappa<-1/5$.

\bibliography{ListeRef_kinetons}

\end{document}